\documentclass[aps,prd,superscriptaddress,nofootnoteinbib]{revtex4-1}
\pdfoutput=1 
\usepackage{amsmath, amsfonts, amsthm, amssymb, graphicx, color, hyperref, mathtools}
\hypersetup{bookmarksnumbered,bookmarksopenlevel=2}

\usepackage{subcaption}
\usepackage[compat=1.0.0]{tikz-feynman}
\usepackage[utf8]{inputenc}
\usepackage{comment}

\allowdisplaybreaks[3]

\def\ba{\begin{eqnarray}}
\def\ea{\end{eqnarray}}
\def\I{{\rm I}}
\def\L{{\rm L}}
\def\J{{\rm J}}
\def\x{{\mathbf x}}

\makeatletter
\def\@fpheader{~}
\preprint{KCL-PH-TH/2018-31 \\\ \\}

\begin{document}

\title{Prospects for axion searches with Advanced LIGO through binary mergers}

\author{Junwu Huang}\email{jhuang@perimeterinstitute.ca}\affiliation{Perimeter Institute for Theoretical Physics, Waterloo, Ontario N2L 2Y5, Canada}
\author{Matthew C. Johnson}\email{mjohnson@perimeterinstitute.ca}\affiliation{Perimeter Institute for Theoretical Physics, Waterloo, Ontario N2L 2Y5, Canada}\affiliation{Department of Physics and Astronomy, York University, Toronto, Ontario, M3J 1P3, Canada}
\author{Laura Sagunski}\email{sagunski@yorku.ca}\affiliation{Perimeter Institute for Theoretical Physics, Waterloo, Ontario N2L 2Y5, Canada}
\affiliation{Department of Physics and Astronomy, York University, Toronto, Ontario, M3J 1P3, Canada}
\author{Mairi Sakellariadou}\email{mairi.sakellariadou@kcl.ac.uk}\affiliation{Theoretical Particle Physics and Cosmology Group, Physics Department, King's College London, University of London, Strand, London WC2R 2LS, UK}
\author{Jun Zhang}\email{jun34@yorku.ca}\affiliation{Perimeter Institute for Theoretical Physics, Waterloo, Ontario N2L 2Y5, Canada}\email{jun34@yorku.ca}
\affiliation{Department of Physics and Astronomy, York University, Toronto, Ontario, M3J 1P3, Canada}

\date{today}
\begin{abstract}

The observation of gravitational waves from a binary neutron star merger by LIGO/VIRGO and the associated electromagnetic counterpart provides a high precision test of orbital dynamics, and therefore a new and sensitive probe of extra forces and new radiative degrees of freedom. Axions are one particularly well-motivated class of extensions to the Standard Model leading to new forces and sources of radiation, which we focus on in this paper. Using an effective field theory (EFT) approach, we calculate the first post-Newtonian corrections to the orbital dynamics, radiated power, and gravitational waveform for binary neutron star mergers in the presence of an axion. This result is applicable to many theories which add an extra massive scalar degree of freedom to General Relativity. We then perform a detailed forecast of the potential for Advanced LIGO to constrain the free parameters of the EFT, and map these to the mass $m_a$ and decay constant $f_a$ of the axion. At design sensitivity, we find that Advanced LIGO can potentially exclude axions with $m_a \lesssim 10^{-11} \ {\rm eV}$ and $f_a \sim (10^{14} - 10^{17}) \ {\rm GeV}$. There are a variety of complementary observational probes over this region of parameter space, including the orbital decay of binary pulsars, black hole superradiance, and laboratory searches. We comment on the synergies between these various observables.
\end{abstract}

\maketitle

\section{Introduction}

The importance of the recent direct detection of gravitational waves from black hole and neutron star mergers can hardly be over-emphasized~\cite{Abbott:2016blz,TheLIGOScientific:2016qqj,Abbott:2016nmj,TheLIGOScientific:2017qsa}. These observations have confirmed the existence of gravitational waves and black holes, among the most important predictions of General Relativity. The binary neutron star event, GW170817, with the coincident electromagnetic observations, has also yielded insight into the nature of short gamma-ray bursts (GRB) and the production of heavy elements in the Universe. What new discoveries might be on the horizon?

It is clear that existing and future gravitational wave (GW) observatories will enable us to learn a great deal about astrophysics~\cite{TheLIGOScientific:2016wfe}. Some of the expected highlights include insight into production mechanisms from population statistics and constraints on the structure of neutron stars from the GW waveform associated with the end stages of inspiral (where tidal effects become important) and the post-merger phase (where a hypermassive neutron star can form). However, the measurement of gravitational waves from binary mergers also provides an unprecedented opportunity to search for fundamental interactions and particles beyond those of the Standard Model of particle physics and General Relativity; see e.g.~\cite{Yunes:2016jcc} for a summary. Several examples include
\begin{itemize}
\item {\bf Self-interactions beyond the Einstein-Hilbert action}\\ It is important to understand how the gravitational sector might be modified by new graviton self-interactions. The possible form of graviton self-interactions is strongly limited by diffeomorphism invariance, as well as causality and analyticity arguments~\cite{Adams:2006sv}. The effect of additional graviton interactions in the Cosmic Microwave Background (e.g.~\cite{Maldacena:2011nz}) and on compact binary mergers (e.g.~\cite{Endlich:2017tqa}) has received some attention in the literature. However, there are a number of challenges associated with the well-posedness of time-evolution in the non-linear theory (e.g.~\cite{Brito:2014ifa,Yunes:2016jcc,Cayuso:2017iqc}) which thus far precludes a full picture of binary mergers. New diffeomorphism invariance breaking graviton interactions can also be introduced to modify binary dynamics as well as gravitational wave propagation. Various forms of massive gravity theories can be tested through the modification of graviton dispersion relations~\cite{deRham:2010kj,Hassan:2011zd,Camanho:2014apa,Hinterbichler:2017qyt,Bonifacio:2017nnt,Rubakov:2004eb,Dubovsky:2004sg}. However, it is currently not known how to calculate the gravitational waveform from binary mergers predicted by massive gravity theories~\cite{deRham:2012fg}. This is because binaries lie in the strongly coupled regime of the theory for any viable value of the graviton mass~\cite{Hinterbichler:2011tt,Dubovsky:2004sg}.

\item {\bf Exotic compact objects}\\ Several proposals exist for compact objects made out of new particles, for example boson stars/axion stars (see Ref.~\cite{Liebling:2012fv} for a review). These new compact objects have masses and sizes (compaction) different from black holes and neutron stars. The measurement of gravitational and electromagnetic radiation resulting from the merger of such objects provides one means of constraining the associated new particles and interactions (see e.g.~\cite{Yunes:2016jcc}).

\item {\bf Light states coupled to gravity}\\ Scalar fields are a ubiquitous feature of physics beyond the Standard Model of particle physics and many extensions of General Relativity. Light scalars that couple to gravity can be probed by black hole superradiance \cite{Arvanitaki:2010sy,Arvanitaki:2014wva,Arvanitaki:2016qwi}. In this case, the large gravitational field in the proximity of black holes and their rapid rotation can source the clustering of large numbers of light bosons, which in turn extract angular momentum from the black hole. Indirect observations of the spin distribution of black hole binaries by Advanced LIGO will shed light on the existence of these light particles~\cite{Arvanitaki:2016qwi}, and searches for continuous wave signals at Advanced LIGO and future gravitational wave detectors might observe these light particles directly~\cite{Arvanitaki:2014wva}.

\item {\bf New force mediator}\\  If coupled to matter, light scalars can mediate new long range interactions between compact objects, commonly termed ``fifth forces" (see \cite{Adelberger:2003zx} and references within). These interactions have been constrained by laboratory experiments \cite{Geraci:2008hb,Kapner:2006si,Hoyle:2004cw} as well as astronomical observations of the solar system (e.g.~\cite{Erickcek:2006vf}) and beyond (e.g.~\cite{Will:2001mx}). Laboratory experiments constrain universally coupled fifth forces to be much weaker than gravitational strength if the force has a range that is longer than a few microns~\cite{Geraci:2008hb,Kapner:2006si,Hoyle:2004cw}. New scalar forces that arise only in a strong gravity or high density environment, however, are unconstrained and can be looked for with Advanced LIGO.

\end{itemize}

In this paper, we focus on this last category, building upon previous work~\cite{Hook:2017psm,Sagunski:2017nzb} suggesting that binary neutron star (NS-NS) and neutron star-black hole (NS-BH) mergers can provide powerful new probes of light scalar force mediators. In particular, we assess the sensitivity of advanced gravitational wave detectors, such as Advanced LIGO and VIRGO, to the effects of axions on the GW waveform in binary mergers. Before proceeding, we review the properties of axions.

The QCD axion was originally introduced as a solution to the strong CP problem \cite{Peccei:1977hh,Peccei:1977ur,Weinberg:1977ma,Wilczek:1977pj}. Experimental searches for a neutron electric dipole moment (EDM) suggest that the strong CP angle is much smaller than $10^{-10}$ \cite{Baker:2006ts}, while CP angles in the Cabibbo-Kobayashi-Maskawa (CKM) matrix have been measured to be $\mathcal{O}(1)$. The puzzling smallness of the strong CP angle can be resolved by introducing the axion particle $a$ with the coupling
\begin{equation}\label{eq:axion_nuclear_coupling}
\frac{a}{f_a} \frac{g_{\rm s}^2}{32 \pi^2} G^{\mu\nu} \tilde{G}_{\mu\nu},
\end{equation}
where $g_{\rm s}$ is the strong coupling constant, $G_{\mu\nu}$ is the gluon field strength with $\tilde{G}_{\mu\nu} = {1\over 2}\epsilon_{\mu\nu\rho\sigma} G^{\rho \sigma}$ its dual, and $f_a$ the axion decay constant. At low energies, the axion field $a$ gets a potential from the coupling to gluons
\begin{equation}
V \approx - m_{\pi}^2 f_{\pi}^2 \sqrt{1 - \frac{4 m_u m_d}{(m_u+m_d)^2}\sin^2 \left(\frac{a}{f_a}\right)},
\end{equation}
where $m_{\pi}$ and $f_\pi$ are respectively the pion mass and decay constant, and $m_{u,d}$ stands for the mass of the up, down quarks. The mass of the QCD axion is related to the axion decay constant by $m_a = 5.7\times 10^{-12} {\, \rm eV}\left(\frac{10^{18} {\,\rm GeV}}{f_a}\right)$, with $m_a \gtrsim 10^{-12} \, {\rm eV}$ if we require $f_a \lesssim m_{\rm Pl}$~\cite{diCortona:2015ldu}. Recently, it was suggested that if the axion sector has a discrete shift symmetry, the potential of the axion can be much shallower, and the axion mass can be exponentially small \cite{Hook:2018jle}, opening up the parameter space over which one should search for a QCD axion. In addition to its coupling to gluons, the QCD axion can have model-dependent couplings to photons and derivative couplings to standard-model fermions (see e.g.~\cite{Agashe:2014kda}). 

There are a number of other motivations for considering pseudo-scalar particles with many of the same properties as the QCD axion, typically referred to as axion-like particles (ALPs). In the following, we generally refer to ALPs as ``axions", which can have any mass $m_a$ and decay constant $f_a$ as well as any subset of the interactions possessed by the QCD axion. For example, string theory compactifications generally predict a number of light axions~\cite{Arvanitaki:2009fg}. Axions might be the dark matter~\cite{diCortona:2015ldu} (or comprise a significant fraction of it) or provide a candidate for dynamical dark energy~\cite{Kamionkowski:2014zda}.

Axions have been constrained by various experiments through their couplings to photons. The axion dark matter experiment (ADMX) published the first constraint on the QCD axion parameter space in the ${\rm \mu eV}$ mass range~\cite{PhysRevLett.120.151301}, assuming that the axion makes up all the dark matter in the Universe. Many experimental searches for axions through their couplings to nucleons, electrons, and photons have recently been proposed to cover a much wider range of masses and coupling strengths~\cite{Sikivie:1985yu,Sikivie:1983ip,Krauss:1985ub,Arvanitaki:2014dfa,Budker:2013hfa,Kahn:2016aff,Baryakhtar:2018doz,Barbieri:2016vwg,TheMADMAXWorkingGroup:2016hpc}. Beside laboratory measurements, indirect measurements of energy loss and energy transport in various astrophysical objects, for instance SN1987 \cite{Chang:2018rso}, have set the most stringent constraint on the QCD axion in the large mass/strong coupling regime. One can also derive constraints on axions from black hole superradiance \cite{Arvanitaki:2014wva}, while for axions with a nuclear coupling one may impose constraints from the measurement of the CP properties of nearby stellar objects \cite{Hook:2017psm}. 

For axions with a nuclear coupling of the form Eq.~(\ref{eq:axion_nuclear_coupling}), it has been shown \cite{Hook:2017psm} that axions can be sourced by compact objects with a high nuclear density, such as neutron stars, thus endowing compact objects with a scalar charge. Such a scalar charge has important implications for NS-NS or NS-BH binary mergers, leading to the emission of axion radiation and an axion-mediated fifth force. Preliminary estimates of these effects on the orbital dynamics and GW waveform were presented in Refs.~\cite{Hook:2017psm,Sagunski:2017nzb}, which demonstrated that in principle there can be a significant and detectable effect to target. 

For theoretical predictions to match the exquisite data quality of GW170817 and future detections, it is necessary to understand the importance of relativistic corrections, which are typically characterized by the post-Newtonian (PN) expansion in $v^2$, the characteristic velocity associated with the orbit. A very useful tool for developing waveforms to arbitrary order in the PN expansion is the EFT framework developed by Goldberger and Rothstein \cite{Goldberger:2004jt,Goldberger:2007hy}. The EFT framework has been used to calculate post-Newtonian corrections to the gravitational potential and quadrupole moments of binary systems, and, as a result, GW waveforms (see \cite{Goldberger:2007hy} for a review), as well as new observable effects beyond General Relativity \cite{Endlich:2017tqa}. One of the merits of an effective field theory approach is that it can be easily extended to include new degrees of freedom and new interactions. In this paper, we extend the effective field theory of gravity for binary systems to include couplings to an axion, and calculate at next-to-leading order (e.g., to 1-PN order) the axion forces between neutron stars as well as axion radiation, both of which are crucial for computing the GW waveform. Our result also applies to theories which include an additional massive scalar degree of freedom coupled to neutron stars. To our knowledge, this result for a {\em massive} scalar does not appear elsewhere in the literature.

A principle additional result of this paper is a forecast demonstrating the potential for Advanced LIGO to look for massive scalars, and in particular axions, with an event similar to GW170817. We find that Advanced LIGO is a very sensitive probe of the scalar charges of neutron stars and the range of the scalar force (or equivalently, the mass of the axion). Translating this into constraints on $\{f_a, m_a \}$ for axions, we find that a GW170817-like event could look for axions in a large region of the theoretically interesting parameter space. This region of parameter space is also the focus of efforts by binary pulsar measurements, black hole superradiance, and laboratory experiments, opening the window for interesting joint analyses. In the optimistic scenario of a detection, these other efforts would provide a means for an independent confirmation of the existence of a new fundamental particle. We hope that this analysis motivates a systematic observational effort to constrain axions with existing and future detections by LIGO-VIRGO, as well as with next-generation gravitational wave detectors.  

The paper is organized as follows. In Section~\ref{Sec:NS}, we summarize the main effects discussed in \cite{Hook:2017psm} and discuss qualitatively the observable consequences of axions on GW waveforms emitted during binary mergers. In Section~\ref{sec:EFT}, we adapt the EFT framework to analytically calculate the corrections to the GW waveform from axion mediated forces and axion radiation, and in Section~\ref{sec:forecast}, we forecast the observable reach of Advanced LIGO at design sensitivity. In Section~\ref{Sec:conclusion}, we conclude and discuss the implications of an Advanced LIGO discovery or exclusion.

Below, we use the conventions: $m_{\rm Pl}^2 = 1/32 \pi G$, $\hbar=c=1$ and $\eta_{\mu\nu} = (1,-1,-1,-1)$.

\section{Neutron stars with axions}\label{Sec:NS}

In this section, we summarize the main properties of axions discussed in~\cite{Hook:2017psm}, which lead to axion mediated forces as well as axion radiation. The coupling of the axion that we search for is the axion nuclear coupling in Eq.~(\ref{eq:axion_nuclear_coupling}). At low energies, and when the axion is the dark matter, this coupling induces an oscillating electric dipole moment of the nucleus, which has been used to look for dark matter axions in the CASPEr experiment~\cite{Budker:2013hfa}. Note, however, that our setup does not require the axion to be the dark matter. It was recently suggested~\cite{Hook:2017psm} that for axions with non-vanishing nuclear coupling, there are corrections to the axion potential when the nucleon number density is non-zero. Neutron stars, and to a lesser extent, white dwarfs and stars, can have large enough nucleon number density to significantly change the shape of the axion potential. Over a wide range of axion parameter space, these corrections can lead to phase transitions in large and dense objects, like neutron stars, implying new constraints on the axion parameter space, and providing new opportunities to look for such axions in Advanced LIGO and future gravitational wave experiments.

The axion becomes tachyonic at $a=0$ inside the neutron star in the region of parameter space where the axion mass in vacuum ($m_a$) and the axion decay constant ($f_a$) satisfy the condition
\begin{equation}
m_a^2 \lesssim  \sigma_{\rm N} n_{\rm N}/4 f_a^2,
\end{equation}
where the parameter $\sigma_N \equiv \sum_{q=u,d} m_q \frac{\partial m_N}{\partial m_q}\approx 59 \,{\rm MeV}$ parametrizes the dependence of the mass of the nucleons on the masses of the quarks, and can be determined from Lattice simulations (see~\cite{Alarcon:2011zs}), while $n_N$ stands for the number density of neutrons inside a neutron star. For axions satisfying the condition
\begin{equation}
R_{\rm NS}  \gtrsim  \frac{1}{w_a}, \, \, \, \,  \, \, \, \, w_a^2 = \frac{\sigma_{\rm N} n_{\rm N}}{4f_a^2} - m_a^2 >0,
\end{equation}
with $R_{\rm NS}$ denoting the radius of the neutron star (NS), the axion is tachyonic inside of the neutron star at the vacuum VEV. This causes the neutron star to develop an axion profile connecting the {\em different} vacua for the axion inside and outside the neutron star. The profile is given approximately by: 
\begin{equation}
  a(r)\simeq\begin{cases}
    \pm \pi f_a\, \,, & \mbox{for}\, \, r<R_{\rm NS} \\
    \pm \pi f_a \frac{R_{\rm NS} \exp [-m_a r]}{r}\, \,, & \mbox{for}\, \, r > R_{\rm NS}.
  \end{cases}
\end{equation}
The axion potential outside the neutron star has minima at $a=0,2 \pi f_a,\ldots$, and therefore this profile connects the inside of the star, where $a = \pm \pi f_a$, to the local minimum of the potential (in vacuum) at $a=0$. 
The axion profiles of neutron stars in a binary interact, leading to changes in the strength of the long range interaction, and therefore modifying the power radiated in gravitational waves. The exact profile will differ slightly due to the density profile of the neutron star and the interaction terms in the axion potential. However, these effects will not be important far from the neutron star interior, which is the relevant regime for our calculation of the inspiral waveform. Before we delve into details of the calculation of the GW waveform both analytically and numerically, we first summarize the main observable effects and where they come from.

\paragraph{Axion mediated forces}

The axion mediates a force between neutron stars when the axion Compton wavelength $ \lambda_a = 1/m_a $ is comparable to, or larger than, the separation between neutron stars. The force is
\begin{equation}\label{AxionForce}
{\bf F} = - \frac{Q_{1}Q_{2}}{4 \pi r^2}\exp[- r/\lambda_a] \, \hat{r},
\end{equation}
at leading order, where $ Q = \pm 4 \pi (\pi f_a R_{\rm NS}) $ are the scalar charges of the neutron stars. Such a force can be either attractive or repulsive, depending on whether the axion field value is the same or opposite sign on the surface of the two neutron stars, respectively. Such a force can be of comparable strength to the gravitational force when the axion decay constant $f_a$ is comparable to the Planck scale $m_{\rm Pl}$. The existence of such a short range interaction can significantly modify the orbital motion of the neutron stars, and therefore the gravitational waveform. At short distances ($r \simeq R_{\rm NS}$), the axion mediated force deviates from the inverse square law due to the induced axion charges and dipole moments of the neutron stars, which can also change the gravitational wave waveform.

\paragraph{Axion radiation} The other major observable effect comes from axion Larmor radiation during the inspiral. The axion radiation turns on when the orbital frequency of the inspiral becomes larger than the mass of the axion. The total power radiated in a neutron star binary inspiral has contributions from both the GR (see e.g.~\cite{Flanagan:1997sx}) and scalar sectors (see e.g.~\cite{PhysRevD.54.1474} for the massless case and~\cite{Alsing:2011er,Hook:2017psm} for the massive case)
\begin{equation}
\frac{dE}{dt } = -\frac{32}{5} G \mu^2 r^4 \Omega^6 -  \frac{1}{4} \frac{\Omega^4 (Q_{ 1} r_1 - Q_{ 2} r_2)^2}{6 \pi} (1-\frac{m_a^2}{\Omega^2})^{3/2} \Theta(\Omega^2 - m_a^2), \label{eq:radaition}
\end{equation}
at leading order, where $\mu = \frac{M_1 M_2}{M_1+M_2}$ is the reduced mass of the system, $\Omega$ is the orbital frequency and $r$ denotes the distance between the two neutron stars. $r_1$ and $r_2$ are the distances from the two neutron stars to the center of mass ($r_1 = r-r_2 = \frac{M_2}{M_1+M_2} r$). The axion radiation is sourced primarily by a time-dependent scalar charge dipole while gravitational radiation is sourced primarily by a time-dependent mass quadrupole. The axion radiation has a weaker frequency dependence when $\Omega \gg m_a$ compared to the gravitational radiation, and therefore it is more important at longer distances compared to gravitational radiation. Observationally, this implies that the GW waveform is altered. In absence of an axion force -- for instance such a force does not exist at leading order in a NS-BH merger -- one gets an additional contribution to $df/dt$ that scales as
\begin{equation}
(df/dt)_{\rm axion} \propto f^{3} \left(1- \frac{m_a}{\pi f}\right)^{3/2} \Theta (\pi f- m_a), 
\end{equation}
at leading order, compared to $df/dt\propto f^{11/3}$ for gravitational radiation.

In the following, we discuss in more detail how to calculate corrections to the gravitational wave waveform due to axion mediated forces and axion radiation, and then present how one can use the observation of binary mergers by Advanced LIGO/Virgo in order to constrain the axion parameter space. 
We consider NS-NS mergers as well as NS-BH mergers and make use of the phenomenological parameters defined below. The charge of the individual compact objects, which determines the size of the axion mediated force, is
\begin{equation}\label{eq:Qtofa}
Q_{1,2} =\begin{cases}  {\pm} 4 \pi^2 f_a R_{\rm NS\,1,\,2}\, \, , &{\rm for \ \ a \ neutron \,\, star} \\
0\, \, ,  & {\rm for \,\, a \,\, black\,\, hole.} \end{cases}
\end{equation}
The dipole moment of the system, which determines the axion radiation, is
\begin{equation}
\vec{P} = {Q_1-Q_2\over 2}\vec{r}_{12},
\end{equation}
where $\vec{r}_{12} = \vec{r}_{1}-\vec{r}_{2}$ is a vector that points from charge $Q_2$ to charge $Q_1$, and $P = |\vec{P}|$ is the magnitude of the dipole moment. In the case of a NS-BH merger, the axion mediated force is zero and the axion radiation is non-zero, while for a NS-NS merger, both the axion mediated force and axion radiation can be present.

\begin{figure}[tbp]
\includegraphics[height=0.3\textwidth]{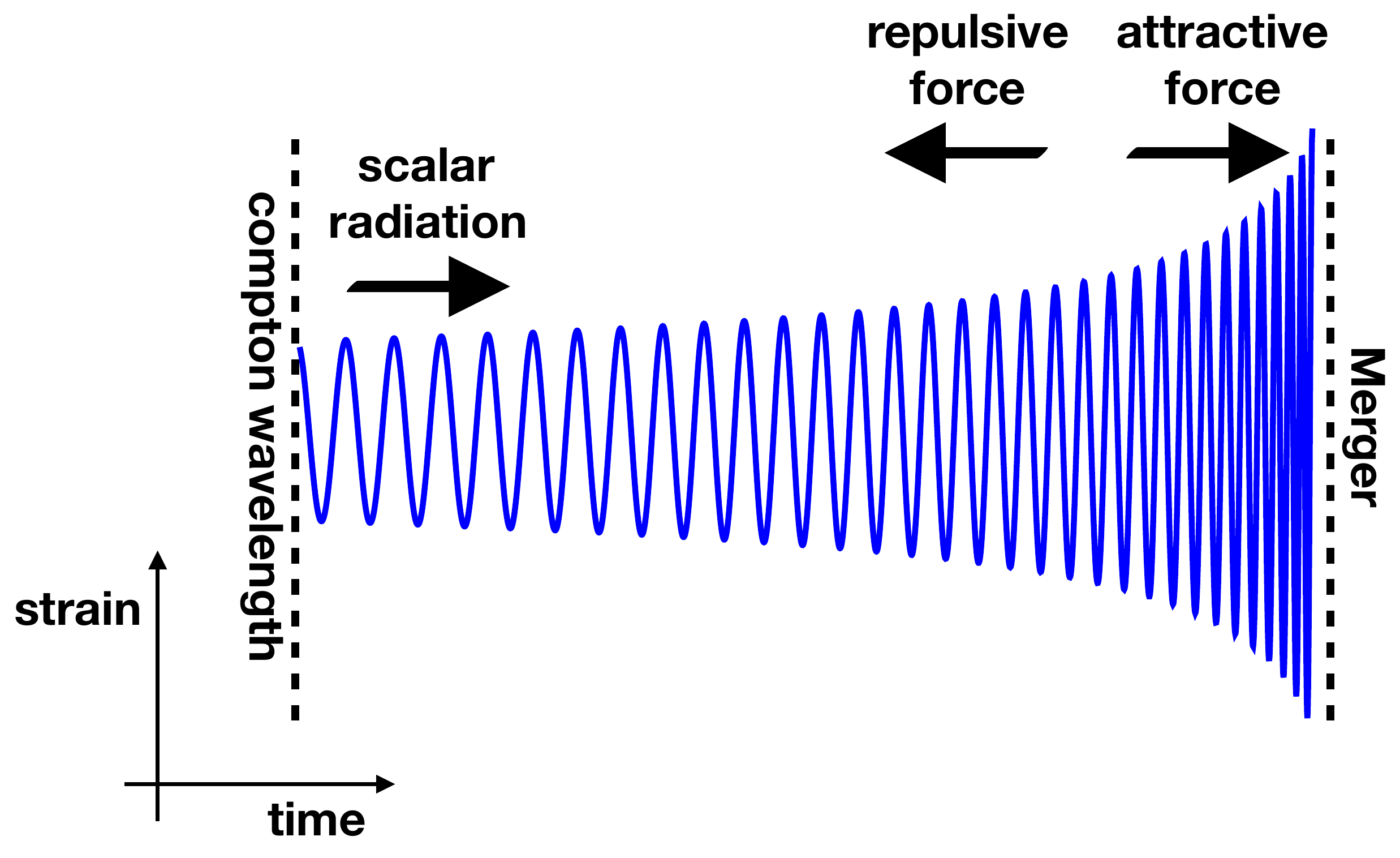} 
\caption{Schematic plot of the strain versus time for a GW waveform emitted during a binary merger in the presence of an axion. The arrows indicate whether the effects of the axion hasten or delay the merger, and therefore shorten or lengthen the chirp (and increase or decrease its pitch, respectively).
} \label{fig:cartoon}
\end{figure}

To gain a qualitative understanding of the effects of axions on the GW waveform, we show a cartoon plot of the strain versus time in Fig.~\ref{fig:cartoon}; a quantitative description can be found below and in Refs.~\cite{Hook:2017psm,Sagunski:2017nzb}. The effect of axions on the waveform is negligible at times before the objects in the binary are separated by roughly a Compton wavelength of the axion. As the orbit decays within the Compton wavelength, scalar radiation can become an important source of orbital energy loss, especially for large Compton wavelengths. This has the effect of increasing the frequency of the GW, and hastening the merger. Scalar radiation is present both for NS-NS and NS-BH binaries. For NS-NS binaries, the effect of the scalar force also becomes important once the orbit has decayed to within the Compton wavelength, and can has a strong effect on the orbital dynamics up to the merger. For neutron stars with the same sign scalar charge, the scalar force is attractive, increasing the frequency of the GW and hastening the merger. For neutron stars of the opposite sign scalar charge, the force is repulsive, decreasing the frequency of the GW, and delaying the merger. In the next section, we discuss these effects in more detail.  

\section{The effect of massive scalars/axions on binary systems}\label{sec:EFT}

In this section, we study the effects of a massive scalar field on the inspiral GW waveform. Our discussion begins with a general scalar field theory, and we then specialize to the axion in Sec.~\ref{sec:matchingsection}. The inspiral dynamics are usually studied using a PN expansion, in which solutions of the Einstein equations are expanded in the characteristic velocity of the system $v$. The inspiral waveform can be obtained to arbitrary accuracy provided the inclusion of sufficiently high order terms. The PN equations of motion can be derived using different methods, all of which lead to the same results at the same PN order. In this paper, we utilize the EFT approach proposed in \cite{Goldberger:2004jt}. We first review the properties of the EFT and then generalize it to include a scalar field.

A neutron star binary simultaneously involves many scales: the size of the neutron star $R_{\rm NS}$, the separation between two neutron stars $r$, and the wavelength of the emitted gravitational waves $\lambda_{\rm GW}$. During the inspiral phase, these three scales have size $R_{\rm NS} \ll r \ll \lambda_{\rm GW}$ and are related to the velocity through $R_{\rm NS}/r \sim r^2/\lambda_{\rm GW}^2\sim v^2 \ll 1$. The smallness of $v$ during the inspiral phase allows us to calculate PN corrections with EFT methods order by order.

To obtain an EFT in the infrared (IR), one can write down an action with all possible terms that respect the symmetries of the system. For example, to calculate the instantaneous potential between binary neutron stars, we represent the neutron stars by two point-like particles, while the mass, spin, and finite size effects of the neutron stars are encoded in the series of couplings between gravitons and the particle world-lines. The value of these couplings can be obtained by utilizing a series of ``matching conditions": comparing the physical quantities, for example, the Newtonian potential, calculated with an EFT approach to the quantities one can directly compute easily in the ultraviolet (UV) limit (e.g., General Relativity).

An {\sl infrared EFT} can also be obtained by ``integrating out" the heavy degrees of freedom in the {\sl ultraviolet  EFT}. Specifically, for a neutron star binary, off-shell gravitons mediating long range interactions between two neutron stars (potential gravitons) typically carry momentum $k \sim 1/r \gg \Omega $, while on-shell gravitons that are emitted by the binary (radiation gravitons) typical carry momentum $k \sim \Omega \sim v/r$ and are therefore  ``lighter" than potential gravitons. The effective action of the low energy radiation graviton can be obtained by integrating out the ``heavy" potential graviton: 
\ba
e^{i S_{\rm eff}\left[\bar{h},\, \x \right]} = \int {\cal D} H_{\mu\nu}\,  e^{i S_{\rm full}\left[H,\, \bar{h},\, \x \right] },
\ea
where $\bar{h}_{\mu\nu}$ denotes the radiation gravitons, $H_{\mu\nu}$ stands for the potential gravitons.

The EFT approach has the advantage of manifesting power counting in the expansion parameter of the theory, which in the case of neutron star binaries is precisely the relative velocity of the binary neutron stars $v$, and therefore makes it easier to track the PN order. As demonstrated in \cite{Goldberger:2004jt}, in the EFT framework, the instantaneous potential as well as the gravitational radiation can be systematically calculated to any order in the PN expansion by including the relevant couplings and ``matching conditions", and working out the corresponding Feynman diagrams. 

In the following, we consider binaries consisting of two scalar charged neutron stars or one scalar charged neutron star and a black hole. Similar to the case of pure gravity, we first write down a series of operators which encode the interactions between the scalar and the members of the binary. In particular, we include operators that characterize the charges and induced dipole moments of the neutron stars. We then calculate the scalar mediated force, and utilize several matching conditions to determine the couplings in the EFT for the axion. We then treat the effects caused by the scalar field perturbatively, and calculate the leading order effects of the scalar field on the 1PN Newtonian potential, as well as on the 1PN gravitational radiation. In the EFT with a scalar, as we demonstrate, we can treat both the scalar charge and the orbital velocity as separate expansion parameters and keep the leading corrections in each.

We consider two scalar charged neutron stars with mass $M_1$ and $M_2$ and charges $Q_1$ and $Q_2$, and with their positions being $\x_1$ and $\x_2$ respectively. As usual, we define
\ba\label{notation}
\mathbf r = \x_1 - \x_2, \quad \mathbf v = \mathbf v_1 - \mathbf v_2, \quad M = M_1+M_2,\quad
\text{and}
\quad \eta = \frac{M_1M_2}{\left(M_1+M_2\right)^2},
\ea
and work in the center of mass frame defined at the corresponding PN order.

\subsection{Binding Energy}\label{sec:bind}

Let us start with the effective action of the binary in pure gravity \cite{Goldberger:2004jt}
\ba\label{SGR}
S_{\rm GR} = - 2 m_{\rm Pl}^2 \int d^4x \sqrt{-g} \left[R -\frac12 \Gamma^\mu \Gamma^\nu g_{\mu
 \nu}\right] -\sum_{n=1,2}  M_n\int d\tau,
\ea
where $\Gamma^\mu = \Gamma^\mu_{\alpha \beta} g^{\alpha \beta}$. The first term is the Einstein-Hilbert action, while the second term fixes the harmonic gauge. The dynamics of the two-body system is described by the third term using the world line approximation. In principle, one could have more generic couplings between gravitons and world lines, which appear at high PN order. Such terms are omitted for the moment.

Now we consider a massive scalar field $\phi$ with
\ba\label{Sphi}
S_{\phi} = \int d^4x \sqrt{-g} \left[\frac12 \partial_\mu \phi \, \partial^\mu \phi - V(\phi)\right].
\ea
We assume a reflection symmetry of $V(\phi)$, as in the axion case, which eliminates couplings such as $\phi^3$ and $\phi h_{00}^2$. Similar to self-interactions of gravitons, self-interaction vertices such as $\phi^4$ and higher powers only contribute at higher order in the PN expansion\footnote{These interactions are only important when the scalar mediated force is much stronger than gravity.}. For these reasons, it is enough to consider $V(\phi) = \frac12 m_{\rm s}^2 \phi^2$. For the charged neutron star solutions discussed in Section~\ref{Sec:NS}, we should consider all possible couplings between the scalar and the world lines that respect the symmetry of the full theory, and therefore the last term in Eq.~(\ref{SGR}) becomes
\begin{equation}\label{eqn:pp}
 S_{\rm pp}=-\sum_{n=1,2}  \int d\tau \left(M_n+ q_n \frac{\phi}{m_{\rm Pl}} + p_n \left(\frac{\phi}{m_{\rm Pl}} \right)^2 + \cdots  \right),
 \end{equation}
where $q_n$ and $p_n$ are the scalar couplings to the neutron star to be determined by utilizing matching conditions. Both $p_i$ and $q_i$ have mass dimension one. Here we only show the terms that contribute up to 1PN. Note that we also do not include $u^{\mu} \partial_{\mu} \phi$ (where $u^{\mu}$ is the 4-velocity), which is proportional to the  equation of motion (up to a total derivative) at leading order, and therefore is a redundant operator. 

To calculate the binding energy as well as radiation in GR, we first expand the metric around Minkowski space 
\ba
g_{\mu\nu} = \eta_{\mu\nu} + \frac{h_{\mu\nu}}{m_{\rm Pl}}.
\ea
Interactions between the point-like particles and the gravitons as well as the scalars are obtained by Taylor expanding action~(\ref{eqn:pp}) in $\mathbf v$. For example, 
\ba
S_{\rm pp} \supset&& M \int d\tau \nonumber \\
=&& M \int dt \left(\frac{1}{2}\mathbf v^2 -\frac{1}{2} \frac{h_{00}}{m_{\rm Pl}}-\frac{h_{0i}}{m_{\rm Pl}} \mathbf v_{i} - \frac{1}{4} \frac{h_{00}}{m_{\rm Pl}} \mathbf v^2 -\frac{1}{2}\frac{h_{ij}}{m_{\rm Pl}}\mathbf v_{i}\mathbf v_{j}+\cdots \right).
\ea
We also have couplings between the scalar field and gravitons from Eq.~(\ref{Sphi}),
\ba
S_{\phi} \supset \int d^4x \, \frac{1}{4m_{\rm Pl}}\left({\mathbf k}\cdot {\mathbf q} - m_{\rm s}^2\right)h_{00}\phi^2  + \frac{1}{4m_{\rm Pl}}\left[\left({\mathbf k}\cdot {\mathbf q} - m_{\rm s}^2\right)\eta^{ij} + 2 {\mathbf k}^i {\mathbf q}^j\right]h_{ij}\phi^2 ,
\ea
where $\eta_{ij} = -\delta_{ij}$ and the dot product between momenta ${\mathbf k}$ and ${\mathbf q}$ is defined as ${\mathbf k}\cdot {\mathbf q} = \delta_{ij}k^i q^j$. Furthermore, we decompose $h_{\mu\nu} = H_{\mu\nu}+\overline{h}_{\mu\nu}$ as well as $\phi = \Phi + \bar{\phi}$ such that $H_{\mu\nu}$ ($\Phi$) represents the off-shell potential graviton (scalar), while $\bar{h}_{\mu\nu}$ ($\bar{\phi}$) is the long-wavelength radiation graviton (scalar). The graviton propagator, which stems from the expansion of the Einstein-Hilbert action with gauge fixing conditions, is given by:
\begin{equation}
\left\langle H_{{\mathbf k} \mu\nu}(x_0) H_{{\mathbf q} \alpha\beta}(0)\right\rangle = - (2\pi)^3 \delta (\mathbf {k+q})\frac{i}{\mathbf k^2} \delta(x_0) P_{\mu\nu,\alpha\beta},
\end{equation}
where $P_{\mu\nu,\alpha\beta} = \frac{1}{2} \left( \eta_{\mu\alpha}\eta_{\nu\beta} + \eta_{\nu\alpha}\eta_{\mu\beta} - \eta_{\mu\nu} \eta_{\alpha\beta}\right)$. Given $P_{00,00} = 1/2$ and $P_{00,ij} = -\eta_{ij}/2$, we have the $H_{00}\phi^2$-vertex
\ba
\begin{gathered}
\begin{tikzpicture}
        \begin{feynman}  
             \vertex(q1);
             \vertex[below=3em of q1](q2);
             \vertex[below=1.5em of q1](i0);
             \vertex[right=1.5em of i0](i);
             \vertex[right= 3em of i](m1);
             \tikzfeynmanset{every vertex={dot}}
               \vertex[right=1.5em of i0](i);
          \diagram*{
          (q1)--[scalar, edge label=\(k\)] (i)
           (q2)--[scalar, edge label'=\(k'\)] (i)
            (m1)--[photon, edge label'=\(q\)] (i)
           };
          \end{feynman}
        \end{tikzpicture}
\end{gathered}=\frac{1}{4m_{\rm Pl}} \frac{-m_{\rm s}^2}{{\mathbf q}^2\left({\mathbf k}^2+m_{\rm s}^2\right)\left({\mathbf k}'^2+m_{\rm s}^2\right)} (2\pi)^3\delta^3({\mathbf k}+{\mathbf k}'+{\mathbf q}).
\ea
Using the power counting rules in \cite{Goldberger:2004jt}, we can find the Feynman diagrams, as shown in Fig.~\ref{fig:LO}, Fig.~\ref{fig:Gv2},  Fig.~\ref{fig:G2}, and Fig.~\ref{fig:3G2}, that contribute to the binding energy up to 1PN order. Among these diagrams, Fig.~\ref{fig:LO2},  Fig.~\ref{fig:Gv22}, Fig.~\ref{fig:Gv23},  Fig.~\ref{fig:G23} and Fig.~\ref{fig:3G23} represent the binding energy from the GR sector \cite{Goldberger:2004jt}. Together with the kinetic term, they give the Lagrangian for the binary in pure gravity:
\ba
L_{\rm GR}= \frac12 \sum_{i=1,2} M_i \mathbf v_i^2 + \frac{GM_1M_2}{r} + L_{\rm EIH},
\ea
with
\ba
L_{\rm EIH}&=& \frac18 \sum_{i=1,2} M_i \mathbf v_i^4  \\ \nonumber
&+& \frac{G M_1M_2}{2r}\left[3\left(\mathbf v_1^2+\mathbf v_2^2\right) - 7 \left(\mathbf v_1 \cdot \mathbf v_2\right) - \frac{\left(\mathbf v_1 \cdot \mathbf r \right)\left(\mathbf v_2 \cdot \mathbf r\right)}{r^2}\right]- \frac{G^2 M_1M_2 (M_1+M_2)}{2 r^2},
\ea
being the Einstein-Infeld-Hoffmann Lagrangian \cite{Einstein:1938yz}. 

\begin{figure}
        \centering
        \begin{subfigure}[b]{0.3\textwidth}
        \centering
        \begin{tikzpicture}
        \begin{feynman}  
             \vertex(i1);
             \vertex[below=3em of i1](i2);
             \vertex[right=3cm of i1](o1);
             \vertex[right=3cm of i2](o2);
             \tikzfeynmanset{every vertex={dot}}
              \vertex[right=1.5cm of i1](v1);
              \vertex[right=1.5cm of i2](v2);
          \diagram*{
            {[edges=plain,fermion]
           (i1) -- (v1) -- (o1),
           (i2) -- (v2) -- (o2),
           },
          (v1)--[scalar, edge label=\(\Phi\)] (v2)
           };
          \end{feynman}
        \draw [line width=0.5mm] (i1|-o1) -- (i1-|o1); 
        \draw [line width=0.5mm] (i2|-o2) -- (i2-|o2); 
        \end{tikzpicture}
        \caption{}\label{fig:LO1}
        \end{subfigure}
        \begin{subfigure}[b]{0.3\textwidth}
        \centering
        \begin{tikzpicture}
        \begin{feynman}  
             \vertex(i1);
             \vertex[below=3em of i1](i2);
             \vertex[right=3cm of i1](o1);
             \vertex[right=3cm of i2](o2);
             \tikzfeynmanset{every vertex={dot}}
              \vertex[right=1.5cm of i1](v1);
              \vertex[right=1.5cm of i2](v2);
          \diagram*{
            {[edges=plain,fermion]
           (i1) -- (v1) -- (o1),
           (i2) -- (v2) -- (o2),
           },
          (v1)--[photon, edge label=\(H_{00}\)] (v2)
           };
          \end{feynman}
        \draw [line width=0.5mm] (i1|-o1) -- (i1-|o1); 
        \draw [line width=0.5mm] (i2|-o2) -- (i2-|o2); 
        \end{tikzpicture}
        \caption{}\label{fig:LO2}
        \end{subfigure}
\caption{Leading order diagrams. In the diagrams above and in Fig.~\ref{fig:Gv2}-Fig.~\ref{fig:3G2} below, the solid black lines are the geodesics of the neutron stars; the dashed lines represent the propagator of the scalar field, and the wiggly lines are the propagator of the graviton.}\label{fig:LO}
\end{figure}
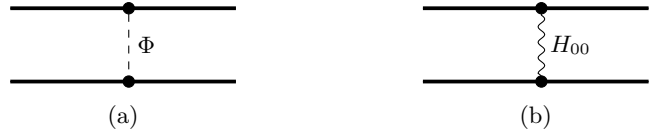
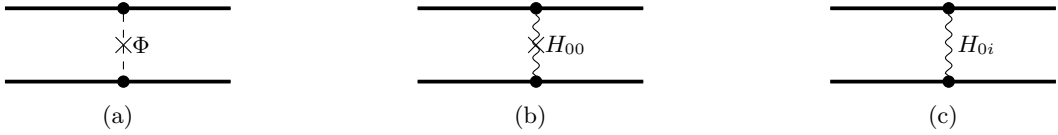
\begin{figure}
        \centering
        \begin{subfigure}[b]{0.3\textwidth}
        \centering
        \begin{tikzpicture}
        \begin{feynman}  
             \vertex(i1);
             \vertex[below=3em of i1](i2);
             \vertex[right=3cm of i1](o1);
             \vertex[right=3cm of i2](o2);
             \tikzfeynmanset{every vertex={dot}}
              \vertex[right=1.5cm of i1](v1);
              \vertex[right=1.5cm of i2](v2);
          \diagram*{
            {[edges=plain,fermion]
           (i1) -- (v1) -- (o1),
           (i2) -- (v2) -- (o2),
           },
          (v1)--[scalar, insertion=0.5,  edge label=\(\Phi\)] (v2)
           };
          \end{feynman}
        \draw [line width=0.5mm] (i1|-o1) -- (i1-|o1); 
        \draw [line width=0.5mm] (i2|-o2) -- (i2-|o2); 
        \end{tikzpicture}
        \caption{}\label{fig:Gv21}
        \end{subfigure}
        \begin{subfigure}[b]{0.3\textwidth}
        \centering
        \begin{tikzpicture}
        \begin{feynman}  
             \vertex(i1);
             \vertex[below=3em of i1](i2);
             \vertex[right=3cm of i1](o1);
             \vertex[right=3cm of i2](o2);
             \tikzfeynmanset{every vertex={dot}}
              \vertex[right=1.5cm of i1](v1);
              \vertex[right=1.5cm of i2](v2);
          \diagram*{
            {[edges=plain,fermion]
           (i1) -- (v1) -- (o1),
           (i2) -- (v2) -- (o2),
           },
          (v1)--[photon, insertion=0.5,  edge label=\(H_{00}\)] (v2)
           };
          \end{feynman}
        \draw [line width=0.5mm] (i1|-o1) -- (i1-|o1); 
        \draw [line width=0.5mm] (i2|-o2) -- (i2-|o2); 
        \end{tikzpicture}
        \caption{}\label{fig:Gv22}
        \end{subfigure}
        \begin{subfigure}[b]{0.3\textwidth}
        \centering
        \begin{tikzpicture}
        \begin{feynman}  
             \vertex(i1);
             \vertex[below=3em of i1](i2);
             \vertex[right=3cm of i1](o1);
             \vertex[right=3cm of i2](o2);
             \tikzfeynmanset{every vertex={dot}}
              \vertex[right=1.5cm of i1](v1);
              \vertex[right=1.5cm of i2](v2);
          \diagram*{
            {[edges=plain,fermion]
           (i1) -- (v1) -- (o1),
           (i2) -- (v2) -- (o2),
           },
          (v1)--[photon, edge label=\(H_{0i}\)] (v2)
           };
          \end{feynman}
        \draw [line width=0.5mm] (i1|-o1) -- (i1-|o1); 
        \draw [line width=0.5mm] (i2|-o2) -- (i2-|o2); 
        \end{tikzpicture}
        \caption{}\label{fig:Gv23}
        \end{subfigure}
\caption{1PN diagrams proportional to $G {\mathbf v}^2$. See Fig.~\ref{fig:LO} for a description of the diagrammatic representation. The crosses in the diagrams above represent the insertion caused by the PN expansion of the propagator.}\label{fig:Gv2}
\end{figure}
\begin{figure}
        \centering
        \begin{subfigure}[b]{0.3\textwidth}
        \centering
        \begin{tikzpicture}
        \begin{feynman}  
             \vertex(i1);
             \vertex[below=3em of i1](i2);
             \vertex[right=3cm of i1](o1);
             \vertex[right=3cm of i2](o2);
             \tikzfeynmanset{every vertex={dot}}
              \vertex[right=1.5cm of i1](v1);
              \vertex[right=1cm of i2](v21);
               \vertex[right= 2cm of i2](v22);
          \diagram*{
            {[edges=plain,fermion]
           (i1) -- (v1) -- (o1),
           (i2) -- (v21)--(v22) -- (o2),
           },
          (v1)--[scalar] (v21),
          (v1)--[scalar] (v22),
           };
          \end{feynman}
        \draw [line width=0.5mm] (i1|-o1) -- (i1-|o1); 
        \draw [line width=0.5mm] (i2|-o2) -- (i2-|o2); 
        \end{tikzpicture}
        \caption{}\label{fig:G21}
        \end{subfigure}
        \begin{subfigure}[b]{0.3\textwidth}
        \centering
        \begin{tikzpicture}
        \begin{feynman}  
             \vertex(i1);
             \vertex[below=3em of i1](i2);
             \vertex[right=3cm of i1](o1);
             \vertex[right=3cm of i2](o2);
             \tikzfeynmanset{every vertex={dot}}
              \vertex[right=1.5cm of i1](v1);
              \vertex[right=1cm of i2](v21);
               \vertex[right= 2cm of i2](v22);
          \diagram*{
            {[edges=plain,fermion]
           (i1) -- (v1) -- (o1),
           (i2) -- (v21)--(v22) -- (o2),
           },
          (v1)--[scalar,  edge label'=\(\Phi\)] (v21),
          (v1)--[photon,  edge label=\(H_{00}\)] (v22),
           };
          \end{feynman}
        \draw [line width=0.5mm] (i1|-o1) -- (i1-|o1); 
        \draw [line width=0.5mm] (i2|-o2) -- (i2-|o2); 
        \end{tikzpicture}
        \caption{}\label{fig:G22}
        \end{subfigure}
        \begin{subfigure}[b]{0.3\textwidth}
        \centering
        \begin{tikzpicture}
        \begin{feynman}  
             \vertex(i1);
             \vertex[below=3em of i1](i2);
             \vertex[right=3cm of i1](o1);
             \vertex[right=3cm of i2](o2);
             \tikzfeynmanset{every vertex={dot}}
              \vertex[right=1.5cm of i1](v1);
              \vertex[right=1cm of i2](v21);
               \vertex[right= 2cm of i2](v22);
          \diagram*{
            {[edges=plain,fermion]
           (i1) -- (v1) -- (o1),
           (i2) -- (v21)--(v22) -- (o2),
           },
          (v1)--[photon,  edge label'=\(H_{00}\)] (v21),
          (v1)--[photon,  edge label=\(H_{00}\)] (v22),
           };
          \end{feynman}
        \draw [line width=0.5mm] (i1|-o1) -- (i1-|o1); 
        \draw [line width=0.5mm] (i2|-o2) -- (i2-|o2); 
        \end{tikzpicture}
        \caption{}\label{fig:G23}
        \end{subfigure}
\caption{1PN diagrams proportional to $G^2$.}\label{fig:G2}
\end{figure}
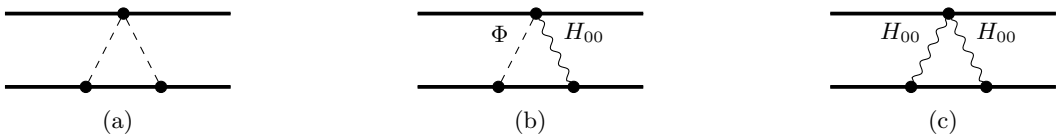
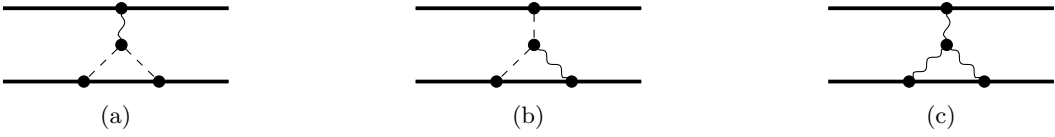
\begin{figure}[h!]
        \centering
        \begin{subfigure}[b]{0.3\textwidth}
        \centering
        \begin{tikzpicture}
        \begin{feynman}  
             \vertex(i1);
             \vertex[below=3em of i1](i2);
             \vertex[right=3cm of i1](o1);
             \vertex[right=3cm of i2](o2);
             \tikzfeynmanset{every vertex={dot}}
              \vertex[right=1.5cm of i1](v1);
              \vertex[below=1.5em of v1](v3);
              \vertex[right=1cm of i2](v21);
              \vertex[right= 2cm of i2](v22);
          \diagram*{
            {[edges=plain,fermion]
           (i1) -- (v1) -- (o1),
           (i2) -- (v21)--(v22) -- (o2),
           },
          (v1)--[photon] (v3),
          (v3)--[scalar] (v21),
          (v3)--[scalar] (v22),
           };
          \end{feynman}
        \draw [line width=0.5mm] (i1|-o1) -- (i1-|o1); 
        \draw [line width=0.5mm] (i2|-o2) -- (i2-|o2); 
        \end{tikzpicture}
        \caption{}\label{fig:3G21}
        \end{subfigure}
        \begin{subfigure}[b]{0.3\textwidth}
        \centering
        \begin{tikzpicture}
        \begin{feynman}  
             \vertex(i1);
             \vertex[below=3em of i1](i2);
             \vertex[right=3cm of i1](o1);
             \vertex[right=3cm of i2](o2);
             \tikzfeynmanset{every vertex={dot}}
              \vertex[right=1.5cm of i1](v1);
              \vertex[below=1.5em of v1](v3);
              \vertex[right=1cm of i2](v21);
              \vertex[right= 2cm of i2](v22);
          \diagram*{
            {[edges=plain,fermion]
           (i1) -- (v1) -- (o1),
           (i2) -- (v21)--(v22) -- (o2),
           },
          (v1)--[scalar] (v3),
          (v3)--[scalar] (v21),
          (v3)--[photon] (v22),
           };
          \end{feynman}
        \draw [line width=0.5mm] (i1|-o1) -- (i1-|o1); 
        \draw [line width=0.5mm] (i2|-o2) -- (i2-|o2); 
        \end{tikzpicture}
        \caption{}\label{fig:3G22}
        \end{subfigure}
        \begin{subfigure}[b]{0.3\textwidth}
        \centering
        \begin{tikzpicture}
        \begin{feynman}  
             \vertex(i1);
             \vertex[below=3em of i1](i2);
             \vertex[right=3cm of i1](o1);
             \vertex[right=3cm of i2](o2);
             \tikzfeynmanset{every vertex={dot}}
              \vertex[right=1.5cm of i1](v1);
              \vertex[below=1.5em of v1](v3);
              \vertex[right=1cm of i2](v21);
              \vertex[right= 2cm of i2](v22);
          \diagram*{
            {[edges=plain,fermion]
           (i1) -- (v1) -- (o1),
           (i2) -- (v21)--(v22) -- (o2),
           },
          (v1)--[photon] (v3),
          (v3)--[photon] (v21),
          (v3)--[photon] (v22),
           };
          \end{feynman}
        \draw [line width=0.5mm] (i1|-o1) -- (i1-|o1); 
        \draw [line width=0.5mm] (i2|-o2) -- (i2-|o2); 
        \end{tikzpicture}
        \caption{}\label{fig:3G23}
        \end{subfigure}
\caption{1PN diagrams with 3-vertices.}\label{fig:3G2}
\end{figure}

Corrections from the scalar field are represented by Fig.~\ref{fig:LO1}, Fig.~\ref{fig:Gv21}, Fig.~\ref{fig:G21}, Fig.~\ref{fig:G22}, Fig.~\ref{fig:3G21}, and Fig.~\ref{fig:3G22}. At 0PN order, the presence of the scalar field leads to an extra diagram
\ba
{\rm Fig.~\ref{fig:LO1}} = i \int dt \frac{q_1 q_2}{m_{\rm Pl}^2} \frac{e^{-m_{s} r}}{4\pi r},
\ea
which contributes a Yukawa potential. At 1PN order, the corrections are given by
\ba
{\rm Fig.~\ref{fig:Gv21}} =  -i \int dt  \frac{q_1 q_2}{8\pi m_{\rm Pl}^2} \frac{e^{-m_{\rm s}r}}{r} \left[ \frac{\left(\mathbf v_1 \cdot \mathbf r\right)\left( \mathbf v_2 \cdot \mathbf r\right)}{r^2}(1+m_{\rm s}r) - \left(\mathbf v_1\cdot \mathbf v_2\right)\right],
\ea
\ba
{\rm Fig.~\ref{fig:G22}} = - i \int dt  \frac{q_1 M_2 q_2}{128\pi^2 m_{\rm Pl}^4} \frac{e^{-m_{\rm s}r}}{r^2} + (1 \leftrightarrow 2),
\ea
\ba
{\rm Fig.~\ref{fig:G21}} = - i \int dt  \frac{p_1 q_2^2}{8\pi^2 m_{\rm Pl}^4} \frac{e^{-2m_{\rm s}r}}{r^2} + (1 \leftrightarrow 2),
\ea
\ba
{\rm Fig.~\ref{fig:3G21}} = i \int dt \frac{M_1 q_2^2}{512\pi^2 m_{\rm Pl}^4} \left[\frac{m_{\rm s}}{r} - \frac{m_{\rm s}}{r} e^{-2m_{\rm s}r} - 2 m_{\rm s}^2 \,{\rm Ei}\left(-2m_{\rm s}r\right) \right] + (1 \leftrightarrow 2),\qquad
\label{eq:potetnial3G21}
\ea
\ba
{\rm Fig.~\ref{fig:3G22}} =  i \int dt \frac{q_1 q_2 M_1} {64\pi^2m_{\rm Pl}^4} \frac{m_{\rm s}}{r}\,  {\cal I}\left(m_{\rm s}r\right) + (1 \leftrightarrow 2),
\ea
where ${\rm Ei}(x) = -\int_{-x}^{\infty} dt\, e^{-t}/t$ is the exponential integral and ${\cal I}(x)$ is a finite integral defined as
\ba
{\cal I}(x) \equiv && \frac{2}{\pi}  \int_0^{\infty}\frac{dk}{k^2+1} \sin\left(k x\right)\arctan k
\ea
The first term in Eq.~(\ref{eq:potetnial3G21}), $\frac{m_{\rm s}}{r}$, comes from the renormalization of the mass of the neutron star from axion mediated interactions at one loop, and can therefore be absorbed by redefining the mass of the neutron star. In the following, we neglect this term in the axion potential since it is not observable. Collecting all the terms gives us the effective Lagrangian from the scalar sector up to 1PN order:
\ba
L_{\phi} =&& 8 G q_1 q_2 
 \frac{e^{-m_{\rm s} r}}{r}
\left[1 - \frac{G(M_1+M_2)}{r}- \frac12 \frac{\left(\mathbf v_1 \cdot \mathbf r\right) \left(\mathbf v_2 \cdot \mathbf r\right)}{r^2}(1+m_{\rm s}r) \right.  
\nonumber \\
&& \qquad \qquad \qquad \;\; \left.+ \frac12 \left(\mathbf v_1\cdot \mathbf v_2\right) - 16G \left(q_1 \frac{p_2}{q_2}+q_2 \frac{p_1}{q_1}\right) \frac{e^{-m_{\rm s}r}}{r} \right] \nonumber \\
 -&& \frac{2 G^2(M_1 q_2^2+ M_2 q_1^2)}{r} m_s\left[ e^{-2m_{\rm s} r}  + 2 m_{\rm s} r\, {\rm Ei}(-2m_{\rm s} r)\right] \nonumber \\
 +&&\frac{16G^2 q_1q_2 (M_1+M_2)}{r} m_{\rm s}\, {\cal I}(m_{\rm s} r).
\ea
For simplicity, we define the following dimensionless parameters
\ba\label{parameters}
q = \frac{q_1q_2}{M^2 \eta}, \quad \alpha^2 = q(\frac{q_1}{M_1} + \frac{q_2}{M_2})^{-2}, \quad
\lambda = \frac{1}{G M m_{\rm s}}, \quad p=\frac{1}{M}\left(q_1 \frac{p_2}{q_2}+q_2 \frac{p_1}{q_2}\right),
\ea
where $-1\le \alpha \le 1$. Note that $q > 0$ if the scalar force between two neutron stars is attractive, and vice versa if repulsive. We also define:
\begin{equation}\label{eq:dimlessromega}
\tilde{r} \equiv r/GM, \ \ \ \ \tilde{\Omega} \equiv GM \Omega.
\end{equation}
The 1PN correction to the  Newtonian potential is given by $V_{\rm GR}+V_{\phi}$ with
\ba\label{VGR}
V_{\rm GR} = M \eta\left\{ - \frac{1}{\tilde{r}} + \frac{3\left(1-3\eta\right)}{8}v^4 +  \frac{1}{2\tilde{r}}\left[\left(3+\eta\right)v^2 + \eta \dot{\tilde{r}}^2 + \frac{1}{\tilde{r}}\right]\right\},
\ea
and
\ba\label{Vphi}
V_{\phi}=&&
-8q M \eta \frac{e^{- \tilde{r}/\lambda}}{\tilde{r}} \left[1 - 16p \frac{e^{-\tilde{r}/\lambda}}{\tilde{r}} \right] \nonumber \\
-&&8q M \eta \frac{e^{- \tilde{r}/\lambda}}{\tilde{r}} \left[- \frac{1}{\tilde{r}} - \frac12\eta(1+\frac{\tilde{r}}{\lambda})\dot{\tilde{r}}^2 + \frac12\eta v^2     \right] \nonumber \\
  +&& 2qM\eta \frac{1}{\tilde{r}\lambda}A(\eta,\alpha)  \left[ e^{-2\tilde{r}/\lambda}+2 \frac{ \tilde{r}}{\lambda} {\rm Ei}\left(-2\frac{\tilde{r}}{\lambda}\right)\right] \nonumber \\
 -&&16qM\eta \frac{1}{\tilde{r}\lambda}\, {\cal I}\left(\frac{\tilde{r}}{\lambda}\right),
\ea
where
\ba
A(\eta,\alpha)\equiv\frac{1+\alpha^2+2\alpha\sqrt{1-4\eta}}{1-\alpha^2}.
\ea

\subsection{Radiation Power}\label{sec:radiation}

We now compute the corrections from the scalar field to the gravitational radiation from the binary at 1PN order. Our goal is to get the corrected radiation power, which is a necessary ingredient for calculating the inspiral waveform. The EFT for the radiation gravitons can be obtained by integrating out the potential graviton $H_{\mu\nu}$ and ``potential scalar'' $\Phi$ defined by $\phi = \Phi + \bar{\phi}$:
\ba\label{pathinte}
e^{i S_{\rm eff}\left[\bar{h},\,\bar{\phi},\, \x \right]} = \int{\cal D} H_{\mu\nu} {\cal D} \Phi \,  e^{i S_{\rm full}\left[h,\,\phi,\, \x\right] },
\ea
where $S_{\rm full} = S_{\rm GR}  + S_{\phi} + S_{\rm pp}$.

\subsubsection{Gravitational wave radiation}
Formally, the source term of the effective action of radiation gravitons can be written as
\ba
S_{\rm eff}^{\rm source} = -\frac{1}{2m_{\rm Pl}} \int d^4 x \, T^{\mu\nu}(\x_a, \bar{h}_{\mu\nu},\bar{\phi})\,  \bar{h}_{\mu\nu},
\ea
where $T^{\mu\nu}$ is the pseudo-energy-stress tensor that can be read off from the path integral~(\ref{pathinte}). To manifest the PN order, it is not enough to just keep $T^{\mu\nu}$ at the right PN order; one should also expand $\bar{h}_{\mu\nu}$ to the right PN order, which is achieved by performing multipole expansions around the center of mass \cite{Goldberger:2004jt}. Multipole expansion of actions is discussed in detail in \cite{Ross:2012fc}. Schematically, one can divide $S_{\rm eff}^{\rm source}$ into two parts: the conserved part and the radiation one. The former has $\bar{h}_{00}$, $\bar{h}_{0i}$ and their spatial derivatives coupled with conserved quantities, such as the ADM mass and momentum, and therefore does not radiate. The latter one has the form
\ba
S_{\rm eff}^{\rm rad} = \int dt \left[\frac12 \I_{g}^{ij} \,R_{0i0j} + \frac16 \I_{g}^{ijk} \, \partial_i R_{0j0k}  + \cdots \right] - \int dt \left[ \frac{1}{3} \epsilon_{imn}  \J_g^{ij}\, R_{0jmn} + \cdots \right],
\ea
where $R_{\mu\nu\sigma\rho}$ is the linearized Riemann tensor defined by the metric $\bar{g}_{\mu\nu} = \eta_{\mu\nu}+\bar{h}_{\mu\nu}/m_{\rm Pl}$ and $\epsilon_{ijk}$ is the Levi-Civita symbol. The $\I_g^{ij}$, $\I_g^{ijk}$, and $\J_g^{ij}$ are the mass quadrupole, mass octupole and current quadrupole, respectively, which, after extensive use of the Ward identity, doing integration by parts, and using the wave equation, are related to the pseudo-energy-stress tensor though
\ba\label{moment}
\I_g^{ij} &=& \int d^3 \x \left(T^{00} + T^{kk} - \frac43 \dot{T}^{0k}\x^k + \frac{11}{42}\ddot{T}^{00}\x^2\right)\left[\x^i\x^j\right]^{\rm STF} + \cdots \\
 \I_g^{ijk} &=& \int d^3 \x \left(T^{00} + T^{ll}\right) \left[\x^i\x^j \x^k\right]^{\rm STF} +\cdots \\
 \J_g^{ij} &=& -\frac12 \int d^3x \left(\epsilon^{ikl}T^{0k}\x^i\x^j + \epsilon^{jkl}T^{0k}\x^i\x^l\right) +\cdots,
\ea
where the dots denote time derivatives, and $[\cdots]^{\rm STF}$ denotes the symmetric trace free components. Note that in the above equations, we have omitted terms that contribute at order higher than 1PN. (We refer the reader to \cite{Ross:2012fc} for more complete and compact expressions.) Finally, the power of gravitational radiation can be calculated using the optical theorem \cite{Goldberger:2004jt},
\ba\label{GWPower}
P_{\rm g} = \frac{G}{\pi T} \int_0^{\omega} d\omega \left[\frac{\omega^6}{5} \left|\I^{ij}\left(\omega\right)\right|^2 + \frac{16\omega^6}{45} \left|\J^{ij}\left(\omega\right)\right|^2  + \frac{\omega^8}{189} \left|\I^{ijk}\left(\omega\right)\right|^2 + \cdots \right].
\ea 
Now, let us get back to the path integral~(\ref{pathinte}) and find the expression for $T^{\mu\nu}$. We only need to calculate $T^{\mu\nu}$ to finite PN order. According to $S_{\rm pp}$, we have $T^{00} \sim v T^{0i} \sim v^2 T^{ij}$ at leading order. On the other hand, we have $r \partial_i R_{\mu\nu\rho\sigma} \sim v R_{\mu\nu\rho\sigma}$, since radiation gravitons carry a typical momentum $k \sim v/r$. With these power counting rules, we conclude that $\I^{ij}_{\rm g} \sim v \I^{ijk}_{\rm g} \sim v \J^{ij}_{\rm g}$ at leading order. Thus, at leading order the gravitational radiation is simply
\ba
\I_{g}^{ij} &=& \int d^3 \x T^{00} \left[\x^i\x^j\right]^{\rm STF},
\ea
with $T^{00} = \sum_{n=1,2} M_n$. We find that the scalar field has no effect on the  gravitational radiation at leading order. Substituting $\I_g^{ij}$ into Eq.~(\ref{GWPower}), one gets the well-known quadrupole formula 
\begin{eqnarray}
\nonumber
P_{\rm GR} = {G\over 5}\langle \dddot{\I}_{ij} \dddot{\I}_{ij}\rangle,
\end{eqnarray}
where the brackets denote a time average. 

\begin{figure}
        \centering
        \begin{subfigure}[b]{0.3\textwidth}
        \centering
        \begin{tikzpicture}
        \begin{feynman} 
             \vertex(i1);
             \vertex[above right=3em of v1](r);
             \vertex[below=3em of i1](i2);
             \vertex[right=3cm of i1](o1);
             \vertex[right=3cm of i2](o2);
             \tikzfeynmanset{every vertex={dot}}
              \vertex[right=1.5cm of i1](v1);
              \vertex[right=1.5cm of i2](v2);
          \diagram*{
            {[edges=plain,fermion]
           (i1) -- (v1) -- (o1),
           (i2) -- (v2) -- (o2),
           },
          (v1)--[scalar, edge label=\(\Phi\)] (v2)
          (v1)--[photon, edge label=\(\bar{h}_{00}\)](r)
           };
           \draw [line width=0.5mm] (i1|-o1) -- (i1-|o1); 
        \draw [line width=0.5mm] (i2|-o2) -- (i2-|o2); 
         \end{feynman}
        \end{tikzpicture}
        \caption{}\label{fig:Rad1}
        \end{subfigure}
        \begin{subfigure}[b]{0.3\textwidth}
        \centering
        \begin{tikzpicture}
        \begin{feynman} 
             \vertex(i1);
             \vertex[below = 1.5em of v1](temp);
             \vertex[right=1.5 cm of temp](r);
             \vertex[below=3em of i1](i2);
             \vertex[right=3cm of i1](o1);
             \vertex[right=3cm of i2](o2);
             \tikzfeynmanset{every vertex={dot}}
              \vertex[right=1.5cm of i1](v1);
              \vertex[below=1.5em of v1](v3);
              \vertex[right=1.5cm of i2](v2);
          \diagram*{
            {[edges=plain,fermion]
           (i1) -- (v1) -- (o1),
           (i2) -- (v2) -- (o2),
           },
          (v1)--[scalar, edge label'=\(\Phi\)] (v2)
          (v3)--[photon, edge label'=\(\bar{h}_{00}\)](r)
           };
           \draw [line width=0.5mm] (i1|-o1) -- (i1-|o1); 
        \draw [line width=0.5mm] (i2|-o2) -- (i2-|o2); 
         \end{feynman}
        \end{tikzpicture}
        \caption{}\label{fig:Rad2}
        \end{subfigure}
        \begin{subfigure}[b]{0.3\textwidth}
        \centering
        \begin{tikzpicture}
        \begin{feynman} 
             \vertex(i1);
             \vertex[below = 1.5em of v1](temp);
             \vertex[right=1.5 cm of temp](r);
             \vertex[below=3em of i1](i2);
             \vertex[right=3cm of i1](o1);
             \vertex[right=3cm of i2](o2);
             \tikzfeynmanset{every vertex={dot}}
              \vertex[right=1.5cm of i1](v1);
              \vertex[below=1.5em of v1](v3);
              \vertex[right=1.5cm of i2](v2);
          \diagram*{
            {[edges=plain,fermion]
           (i1) -- (v1) -- (o1),
           (i2) -- (v2) -- (o2),
           },
          (v1)--[scalar, edge label'=\(\Phi\)] (v2)
          (v3)--[photon, edge label'=\(\bar{h}_{ij}\)](r)
           };
           \draw [line width=0.5mm] (i1|-o1) -- (i1-|o1); 
        \draw [line width=0.5mm] (i2|-o2) -- (i2-|o2); 
         \end{feynman}
        \end{tikzpicture}
        \caption{}\label{fig:Rad3}
        \end{subfigure}
\caption{Corrections from the scalar field on $T^{00}$ and $T^{kk}$ at 1PN.}\label{fig:Rad}
\end{figure}
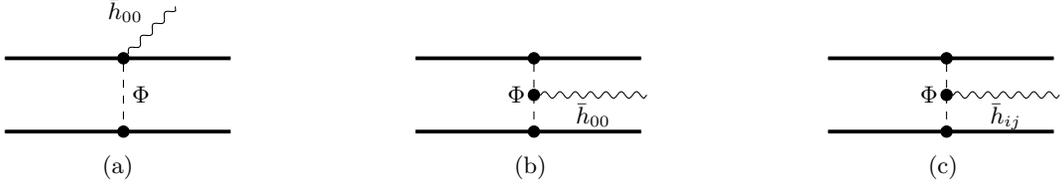

Calculation of the gravitational radiation power to next-to-leading-order needs the leading mass octupole, the leading current quadrupole, and the mass quadrupole up to ${\cal O} \left(v^2\right)$. According to Eq.~(\ref{moment}), we only need to calculate $T^{00}$, $T^{kk}$ and $T^{0i}$ in $\I_g^{ij}$ up to ${\cal O} \left(v^2\right)$. The leading corrections from the scalar field are shown in Fig.~\ref{fig:Rad}, from which we find that all corrections have a magnitude of $q v^2$. For small $q$ (as considered below), they can be simply neglected at 1PN and therefore the gravitational wave radiation power is the same as in the case of pure gravity:
\ba\label{Pg}
P_{\rm g} =\frac{32}{5}G M^2 \eta^2 r^4\Omega^6\left[(1+X)^2 + \frac{19}{21}(1-3\eta)X r^2\Omega^2  + \left(\frac{769}{336}-\frac{2772}{336}\eta\right)r^2\Omega^2 \right].
\ea
with
\ba
X= - (1-2\eta) \frac{GM}{r},
\ea 
where $r$ is related to $\Omega$ through the modified Kepler's law at 1PN. Note that we do not expand Eq.~(\ref{Pg}) in $v^2$ at this point.

\subsubsection{Scalar radiation}

In addition to gravitational radiation, there is scalar radiation in the presence of the scalar field. A scalar field with a Compton wavelength much larger than the binary separation leads to scalar radiation that dominates the energy loss, and therefore is severely constrained by e.g. observations of binary pulsars~\cite{Alsing:2011er}. Similarly to gravitational radiation, the source term for scalar radiation can be written as
\ba\label{Jphi}
S_{\rm eff}^{\rm source} =  \int dt \, \J \,  \bar{\phi}(t,\x),
\ea
where $\J$ is calculated in a PN expansion. In principle, to get the scalar radiation power at 1PN, we need to calculate $\J$ to 2PN order. This is because the power of dipole radiation is usually one PN order lower than that of quadrupole radiation. However, for small scalar charge (as considered below), we only need to calculate $\J$ to 1PN order for dipole scalar radiation and at leading order for quadrupole scalar radiation.

Diagrams that contribute to $\J$ up to 1PN order are shown in Fig.~\ref{fig:ScaRad1}, where
\ba
{\rm Fig.~\ref{fig:ScaRad11}} = -i \sum_{n=1,2} \int dt \, \left(1-\frac{1}{2}{\mathbf v}_n^2\right)\frac{q_n}{m_{\rm Pl}} \bar{\phi}, 
\ea 
\ba
{\rm Fig.~\ref{fig:ScaRad12}} = i  \int dt \, \left(\frac{q_1p_2+q_2p_1}{4\pi m_{\rm Pl}^2} \frac{e^{-m_{\rm s} r}}{r}\right)\frac{\bar{\phi}}{m_{\rm Pl}} , 
\ea 
\ba
{\rm Fig.~\ref{fig:ScaRad13}} = i \int dt \,  \left(\frac{q_1M_2+q_2M_1}{32\pi m_{\rm Pl}^2} \frac{1}{r}\right)\frac{\bar{\phi}}{m_{\rm Pl}} , 
\ea 
and
\ba
{\rm Fig.~\ref{fig:ScaRad14}} = i \int dt \,  \left(\frac{q_1M_2+q_2M_1}{32\pi m_{\rm Pl}^2} \frac{1-e^{-m_{\rm s} r}}{r}\right)\frac{ \bar{\phi}.}{m_{\rm Pl}}
\ea

%
\begin{figure}
        \centering
        \begin{subfigure}[b]{0.2\textwidth}
        \centering
        \begin{tikzpicture}
        \begin{feynman} 
             \vertex(i1);
             \vertex[above right=3em of v1](r);
             \vertex[below=3em of i1](i2);
             \vertex[right=3cm of i1](o1);
             \vertex[right=3cm of i2](o2);
             \tikzfeynmanset{every vertex={dot}}
              \vertex[right=1.5cm of i1](v1);
              \vertex[right=1.5cm of i2](v2);
              \vertex[below=0.75em of v1]{\(\sqrt{1-{\mathbf v}_n^2}\)};
          \diagram*{
            {[edges=plain,fermion]
           (i1) -- (v1) -- (o1),
           (i2) -- (v2) -- (o2),
           },
          (v1)--[scalar, edge label=\(\bar{\phi}\)](r)
           };
         \draw [line width=0.5mm] (i1|-o1) -- (i1-|o1); 
         \draw [line width=0.5mm] (i2|-o2) -- (i2-|o2); 
         \end{feynman}
        \end{tikzpicture}
        \caption{}\label{fig:ScaRad11}
        \end{subfigure}
        \begin{subfigure}[b]{0.2\textwidth}
        \centering
        \begin{tikzpicture}
        \begin{feynman} 
             \vertex(i1);
             \vertex[above right=3em of v1](r);
             \vertex[below=3em of i1](i2);
             \vertex[right=3cm of i1](o1);
             \vertex[right=3cm of i2](o2);
             \tikzfeynmanset{every vertex={dot}}
              \vertex[right=1.5cm of i1](v1);
              \vertex[right=1.5cm of i2](v2);
          \diagram*{
            {[edges=plain,fermion]
           (i1) -- (v1) -- (o1),
           (i2) -- (v2) -- (o2),
           },
          (v1)--[scalar, edge label=\(\Phi\)] (v2)
          (v1)--[scalar, edge label=\(\bar{\phi}\)](r)
           };
           \draw [line width=0.5mm] (i1|-o1) -- (i1-|o1); 
        \draw [line width=0.5mm] (i2|-o2) -- (i2-|o2); 
         \end{feynman}
        \end{tikzpicture}\caption{}\label{fig:ScaRad12}
        \end{subfigure}
        \begin{subfigure}[b]{0.2\textwidth}
        \centering
        \begin{tikzpicture}
        \begin{feynman} 
             \vertex(i1);
             \vertex[above right=3em of v1](r);
             \vertex[below=3em of i1](i2);
             \vertex[right=3cm of i1](o1);
             \vertex[right=3cm of i2](o2);
             \tikzfeynmanset{every vertex={dot}}
              \vertex[right=1.5cm of i1](v1);
              \vertex[right=1.5cm of i2](v2);
          \diagram*{
            {[edges=plain,fermion]
           (i1) -- (v1) -- (o1),
           (i2) -- (v2) -- (o2),
           },
          (v1)--[photon, edge label=\(H_{00}\)] (v2)
          (v1)--[scalar, edge label=\(\bar{\phi}\)](r)
           };
           \draw [line width=0.5mm] (i1|-o1) -- (i1-|o1); 
        \draw [line width=0.5mm] (i2|-o2) -- (i2-|o2); 
         \end{feynman}
        \end{tikzpicture}
        \caption{}\label{fig:ScaRad13}
        \end{subfigure}
        \begin{subfigure}[b]{0.2\textwidth}
        \centering
        \begin{tikzpicture}
        \begin{feynman} 
             \vertex(i1);
             \vertex[below = 1.5em of v1](temp);
             \vertex[right=1.5 cm of temp](r);
             \vertex[below=3em of i1](i2);
             \vertex[right=3cm of i1](o1);
             \vertex[right=3cm of i2](o2);
             \tikzfeynmanset{every vertex={dot}}
              \vertex[right=1.5cm of i1](v1);
              \vertex[below=1.5em of v1](v3);
              \vertex[right=1.5cm of i2](v2);
          \diagram*{
            {[edges=plain,fermion]
           (i1) -- (v1) -- (o1),
           (i2) -- (v2) -- (o2),
           },
          (v1)--[scalar, edge label'=\(\Phi\)] (v3)
          (v3)--[photon, edge label'=\(H_{00}\, H_{ij}\)] (v2)
          (v3)--[scalar, edge label'=\(\bar{\phi}\)](r)
           };
           \draw [line width=0.5mm] (i1|-o1) -- (i1-|o1); 
        \draw [line width=0.5mm] (i2|-o2) -- (i2-|o2); 
         \end{feynman}
        \end{tikzpicture}
        \caption{}\label{fig:ScaRad14}
        \end{subfigure}
\caption{Diagrams contribute to scalar radiation at leading order, $v^2$ and $p v^2$.}\label{fig:ScaRad1}
\end{figure}
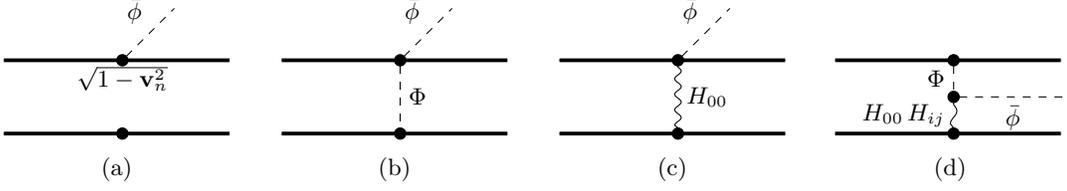
\begin{figure}
        \centering
        \begin{subfigure}[b]{0.3\textwidth}
        \centering
        \begin{tikzpicture}
        \begin{feynman}  
             \vertex(i1);
             \vertex[above right=3em of v1](r);
             \vertex[below=3em of i1](i2);
             \vertex[right=3cm of i1](o1);
             \vertex[right=3cm of i2](o2);
             \tikzfeynmanset{every vertex={dot}}
              \vertex[right=1.5cm of i1](v1);
              \vertex[right=1cm of i2](v21);
               \vertex[right= 2cm of i2](v22);
          \diagram*{
            {[edges=plain,fermion]
           (i1) -- (v1) -- (o1),
           (i2) -- (v21)--(v22) -- (o2),
           },
          (v1)--[scalar,  edge label'=\(\Phi\)] (v21),
          (v1)--[photon,  edge label=\(H_{00}\)] (v22),
          (v1)--[scalar, edge label=\(\bar{\phi}\)](r)
           };
          \end{feynman}
        \draw [line width=0.5mm] (i1|-o1) -- (i1-|o1); 
        \draw [line width=0.5mm] (i2|-o2) -- (i2-|o2); 
        \end{tikzpicture}
        \caption{}\label{fig:ScaRad23}
        \end{subfigure}
\caption{Corrections from the scalar field proportional to $p v^4$.}\label{fig:ScaRad2}
\end{figure}
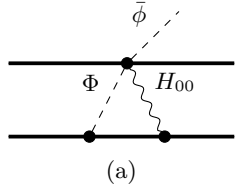
In some cases, one may also want to include terms proportional to $p v^4$, which come from the diagram in Fig.~\ref{fig:ScaRad2} and contribute
\ba
{\rm Fig.~\ref{fig:ScaRad23}} = -i \int dt \,  \left(\frac{q_1M_1p_2+q_2M_2p_1}{128\pi^2 m_{\rm Pl}^4} \frac{e^{-m_{\rm s} r}}{r}\right)\frac{\bar{\phi}}{m_{\rm Pl}} .
\ea
Collecting all the diagrams shown in Fig.~{\ref{fig:ScaRad1}}, we have $J = - \sum_{n} \tilde{q}_n/m_{\rm Pl}$ with
\ba
\tilde{q}_n = q_n \sum_{m\neq n} 1 - \frac12\frac{M_m^2}{M^2} r^2 \Omega^2 - \left(2-e^{-m_{\rm s} r} \right)\frac{GM_m}{r}-\frac{8G p_m}{r} e^{-m_s r}. 
\ea
The radiation power can be calculated using
\ba
P_{\rm s} = \frac{1}{4\pi^2 T}   \sum_{l=0}^{\infty} \frac{1 }{l! (2l+1)!!}  \int  d\omega\, \omega\left(\omega^2-m_{s}^2\right)^{l+1/2}   \left|\I^\L(\omega) \right|^2 ,
\ea 
where the multipole moments $\I^\L$, which arise from multipole expanding the source action~(\ref{Jphi}), are given by \cite{Ross:2012fc}
\ba
\I^\L = \sum_{p=0}^{\infty}\frac{(2l+1)!!}{(2p)!!(2l+2p+1)!!}\, \partial_{t}^{2p} J r^{2p} \, x^\L_{\rm STF}\,.
\ea
Here $L$ denotes a collection of index $i_1 i_2 ... i_l$, and $x^{\L} = x^{i_1}x^{i_2}...x^{i_l}$. For $l=0$, we have $\mathbf x_s^2 \propto r$, therefore $d \I / dt \propto \dot{r}$ which vanishes at 1PN. Thus, there is no monopole scalar radiation for circular orbits.\\
For $l=1$ we obtain,
\ba\label{Ps1}
P_{\rm s}^{l=1} = \frac{1}{12\pi}\frac{(\tilde{q}_1M_2-\tilde{q}_2M_1)^2}{ M^2 m_{\rm Pl}^2} \left(1-\frac{m_{\rm s}^2}{\Omega^2}\right)^{3/2} r^2\Omega^4,
\ea
and $l=2$ yields   
\ba\label{Ps2}
P_{\rm s}^{l=2} =\frac{4}{15\pi}\frac{(q_1M_2^2+q_2M_1^2)^2}{ M^4 m_{\rm Pl}^2} \left(1-\frac{m_{\rm s}^2}{4\Omega^2}\right)^{5/2} r^4\Omega^6.
\ea

\subsection{Matching to Axions}\label{sec:matchingsection}

Consider the axion model of \cite{Hook:2017psm}, which yields scalar charged neutron stars for
\ba\label{condition}
m_a^2 f_a^2 \ll \frac{\sigma_N \rho_{\rm NS}}{4 m_N}.
\ea 
To make use of the results above, we have to fix the parameters $q$ and $p$, defined in Eq.~(\ref{parameters}), by matching with the full theory. According to \cite{Hook:2017psm}, the charged neutron stars have constant axion field value at the edge of the stars. In this case, the scalar potential between two charged neutron stars of radius $R_{\rm NS}^{(1)}$ and $R_{\rm NS}^{(2)}$ can be calculated at Newtonian order using the image charge method; it reads
\ba
V(r) = -\frac{Q_1Q_2}{4\pi r} \left(1 - \frac{R_{\rm NS}^{(1)}+R_{\rm NS}^{(2)}}{2r}\right) + {\mathcal O}\left(\frac{1}{r^3}\right).
\label{eq:V1}
\ea
We can match to the EFT by taking the limit where $m_s \rightarrow 0$ and neglecting the velocity-dependent terms in Eq.~\ref{Vphi}. The finite size effects of the neutron stars, and therefore $q$ and $p$, should not depend on $m_a$, as long as the condition Eq.~(\ref{condition}) is satisfied. We can therefore extend these relations to non-zero mass. In the massless limit, the pure scalar potential between two static sources according to the EFT is
\ba
V_{s}^{m_s \rightarrow 0} (r)= -
 \frac{8 Gq_1q_2}{r} \left(1  -  \frac{16GM p}{r} \right) + {\mathcal O}\left(\frac{1}{r^3}\right).
\label{eq:V2}
\ea
Comparing Eq.~(\ref{eq:V1}) to Eq.~(\ref{eq:V2}), we find
\begin{equation} 
q_i = Q_i m_{\rm Pl} \quad \text{and} \quad p = {R_{\rm NS}^{(1)}+R_{\rm NS}^{(2)}\over 16 G M}.
\label{eq:p1p2}
\end{equation}
Note that despite of the simple relation, $q_i$ and $Q_i$ are different since the former is the Wilson coefficient we introduced in the EFT as well as the free parameter in the waveform, while the latter is the scalar charge of the neutron star in the specific axion model. The parameter $p$ is therefore bounded from below by $1/8$ in the limit $R_{\rm NS}^{(i)} = 2 G M_i$ and resides in the range $(0.25,0.4)$ for neutron stars that are consistent with various constraints (see reference~\cite{Annala:2017llu} and reference within). Such a requirement ensures that corrections to the axion potential and radiation, enhanced by $16 p$ and $8 p$ compared to the corresponding GR corrections for potential and radiation, respectively, are the leading corrections that help distinguish axion mediated interactions from gravity.

In terms of the dimensionless variables of Eq.~\eqref{parameters} and~\eqref{eq:dimlessromega}, the leading corrections from the scalar sector are therefore given by Fig.~\ref{fig:LO1} and Fig.~\ref{fig:G21}:
\ba\label{Vssimple}
V_{a} = -8q M \eta \frac{e^{-\tilde{r}/\lambda}}{\tilde{r}} \left[1 - 16p \frac{e^{-\tilde{r}/\lambda}}{\tilde{r}} \right],\label{eq:Va}
\ea
and the modified Kepler relation is
\ba\label{Keplershort}
\tilde{\Omega}^2 = \frac{1}{\tilde{r}^3} \left[1+\frac{\eta -3}{\tilde{r}} 
+ 8 q \left(1+\frac{\tilde{r}}{\lambda}\right)e^{-\tilde{r}/\lambda} -256qp\left(1+\frac{\tilde{r}}{\lambda}\right) \frac{e^{-2\tilde{r}/\lambda}}{\tilde{r}}\right].
\ea
These constitute a minimal model for the effects of the axion on the binary. 

For future reference, in Tab.~\ref{tab:dimension}, we summarize the dimension and the magnitude of the EFT parameters as they are in the axion neutron star model studied in \cite{Hook:2017psm}.
\begin{table}
\centering
\begin{tabular}{c|c|c|c|c|c|c|c|c}
& $q_n$ & $q$ & $p_n$ & $p$ & $\lambda$ & $\tilde{r}$ & $\tilde{\Omega}$ & $\alpha^2 $\\
\hline
def  & $-$ & $\frac{q_1 q_2}{M^2 \eta}$ & $-$ & $ \frac{q_1 p_2}{q_2 M}+ \frac{q_2 p_1}{q_1 M}$ & $\frac{1}{G M m_{\rm s}}$ & $\frac{r}{GM}$ & $GM \Omega$ & $q \left(\frac{q_1}{M_1}+\frac{q_2}{M_2}\right)^{-2}$ \\
dim  &1 & 0 & 1 & 0 & 0 & 0 & 0 & 0 \\
order & $m_{\rm Pl} R_{\rm NS} f_a$ & $\frac{(R_{\rm NS} f_a)^2}{G M^2}$ & $m_{\rm Pl}^2 R_{\rm NS}$ & $\frac{R_{\rm NS}}{16G M}$ & $\frac{1}{G M m_{\rm s}}$ & $v^{-2}$ & $v^3$ & 1 \\
value & $-$ & $ 0.4 $ & $-$ & $ 1.2 $ & $ 100 $ & $v^{-2}$ & $v^3$ & $-$ \\
\end{tabular}
\caption{In this table, we summarize the definition (def), mass dimension (dim) and rough order of magnitude of the parameters defined in the EFT since they have non-standard dimensions as charges and dipole moments. The three dimensionless parameters $|q| \,(\leq 1)$, $p \,(>1/8)$ and $\lambda$, help us keep track of the orders of perturbative expansion in different regimes. In the last line of the table, we also provide the approximate value of the dimensionless perturbative expansion parameters with a set of benchmark parameters $f_a = 10^{17} \,{\rm GeV}$, $m_{\rm s} = 10^{-12} \,{\rm eV}$, $R_{\rm NS} = 18 \,{\rm km}$, $M_{\rm NS} = 1.25 M_{\odot}$}\label{tab:dimension}
\end{table}

\section{Waveform and prospects for detection with Advanced LIGO}\label{sec:forecast}

In this section, we first calculate NS-NS and NS-BH merger waveforms with axion induced corrections based on the axion mediated force and axion radiation found in the previous section. We then compare this GW waveform to the one within General Relativity, and assess the detectability of these corrections. The result of such a comparison is presented as a projected constraint on the axion parameter space. The method used in this section can be adapted to the study of any other theory where a light massive scalar is coupled to the neutron star or other compact objects. In this section, we only keep the leading corrections to the potential and radiation, to an order that is relevant for breaking the degeneracy between axion induced corrections to the gravitational waveform and post-Newtonian corrections. In principle, one can consider the spin and tidal effects by including higher PN terms. However, as shown in Appendix~\ref{App:A}, including of higher PN terms does not significantly affect the constraints on the EFT parameters. We therefore neglect these effects.

\subsection{Waveform}
The inspiral waveform measured in a gravitational wave detector is of the form of~\cite{Flanagan:1997sx,Allen:2005fk}
\ba\label{waveform}
h(t) = h_0(t) \cos\phi(t),
\ea
where
\ba
h_0(t)=\frac{4 \cal{Q}}{D_L} G M\eta \, \Omega^2 r^2 \quad \text{and} \quad \phi(t) = \int 2\pi f dt,
\ea
with $D_L$ being the luminosity distance to the source and where $\cal{Q}$ encodes the detector response as a function of the angular position and orientation of the binary. For convenience, we neglect the cosmological red-shifting of the observed frequency of gravitational radiation (motivated by the limited horizon for neutron star binary mergers with current interferometers). In addition, we assume an ideally oriented binary and set ${\cal Q}=1$. If $d\ln h_0/dt \ll d\phi/dt$ and $d^2\phi/dt^2 \ll (d\phi/dt)^2$, the Fourier transform of the time-domain waveform
\ba
\tilde{h}(f) \equiv \int_{-\infty}^{\infty} e^{2\pi ift}h(t) dt,
\ea
can be computed using the stationary phase approximation, 
\ba
\tilde{h}(f) \simeq H(f)\, e^{i \Psi(f)},
\ea
where 
\ba
H(f) = \frac12 h_0(t) \left(\frac{df}{dt}\right)^{-1/2}, \quad \text{and} \quad \Psi(f) =  2\pi f t-\phi(f) -\frac{\pi}{4}.
\ea
In the above two equations,  $t$ should be thought as a function of $f$ and defined as the time at which $d\phi/dt = 2\pi f$. Usually one can solve for $r(\Omega)$ from the modified Kepler's law, e.g. Eq.~(\ref{Keplershort}), and then get the analytical frequency domain waveform at 1PN. However, in the presence of a massive scalar field, $\Omega^2$ is not analytical in terms of the PN parameters, and therefore we cannot solve $r(\Omega)$ in general. For this reason, we first calculate $H$ and $\phi$ in terms of $r$, and translate them to $f$ using a numerical interpolation function $r(\Omega)$ when we generate the waveform. The system we solve is given by
\ba
&&\left(\frac{df}{dt}\right)^{-1/2} = \left(-\frac{\pi}{P}\frac{dE}{dr} \frac{dr}{d\Omega}\right)^{1/2}, \\
&&t(r)  =  -\int P(r)^{-1} \left(\frac{dE}{dr}\right) dr,  \\
&&\phi(r)  = -\int 2\Omega(r) P(r)^{-1} \left(\frac{dE}{dr}\right) dr,
\ea
where $E = \frac12 M\eta r^2 \Omega^2 + V_{\rm GR} + V_{a}$, $P=P_{\rm g}+P_{\rm a}$ and we have used $\Omega = \pi f$. Together with Eqs.~(\ref{VGR}), (\ref{Vphi}), (\ref{Pg}), (\ref{Ps1}) and (\ref{Ps2}), we can solve for $\Psi$ and $H$, and eventually get the waveform numerically.

The potential and radiation terms we calculated in sections \ref{sec:bind} and \ref{sec:radiation} contain all corrections at leading order in axion charge, and up to next to leading order in the PN expansion. These corrections are all needed if we were to extract information about axions from LIGO data. The waveform we calculated numerically, and subsequently use to estimate the reach for the axions using a Markov Chain Monte Carlo (MCMC) sampler, however, only take corrections Eq.~(\ref{VGR}), Eq.~(\ref{eq:Va}) for the potential, and Eq.~(\ref{Ps1}), Eq.~(\ref{Ps2}) for the radiation into consideration. The potential terms include the Newtonian and 1PN corrections to the gravitational potential, the leading order axion potential and its leading correction due to ``image charges''. The radiation terms include the leading and next leading order gravitational wave quadrupole, as well as the axion dipole and quadrupole radiation and its leading correction from the induced dipole. These terms are sufficient to break the degeneracy between axion induced corrections to the gravitational waveform and post-Newtonian corrections~\footnote{The axion dipole radiation power has a different frequency dependence  compared to the quadrupole gravitational wave power, and therefore, the leading axion dipole radiation itself breaks the degeneracy between axion radiation and gravitational wave radiation.}. Higher PN corrections on the GR side, while giving rise to qualitatively similar behavior to the scalar sector (e.g. hastening the merger), will not be degenerate with the scalar corrections to the waveform (e.g. because of their different frequency dependence).

\subsection{Forecast}

Given a high signal-to-noise (SNR) detection of a merger event, it is possible to use the measured inspiral waveform not only to infer the parameters of the binary, but also to derive constraints on parameters in the scalar sector: $q_{1,2}$, $p_{1,2}$, and $\lambda$. A measured signal $s(t,\boldsymbol{\bar{\theta}})$ consists of a noise realization $n(t)$ and a merger waveform $\bar{h}(t,\boldsymbol{\bar{\theta}})$ depending on the ``true" parameters $\boldsymbol{\bar{\theta}}$, namely $s(t,\boldsymbol{\bar{\theta}}) = n(t) + \bar{h}(t,\boldsymbol{\bar{\theta}})$. For a set of template waveforms $g(t,\boldsymbol{\theta})$, which depend on a set of candidate parameters $\boldsymbol{\theta}$, the likelihood function is 
\begin{equation}
\mathcal{L} (s | \boldsymbol{\theta}) = \mathcal{N} \exp\left[ -\frac{1}{2} \left(\,s - g \,|\,s-g\,\right) \right]\,, 
\end{equation}
where $\mathcal{N}$ is a normalization factor \cite{Finn:1992wt}. Given two signals $h(t)$ and $g(t)$, the inner product $\left(\,h\,|\,g\,\right)$ on the vector space of signals is defined as
\ba
\left(\,h\,|\,g\,\right) = 2\int_0^{\infty} \frac{\tilde{h}^*(f) \tilde{g}(f) + \tilde{h}(f)\tilde{g}^*(f)}{S_n(f)}df\,,
\ea
where $S_n(f)$ is the detector noise spectral density and $\tilde{h}, \tilde{g}$ are the Fourier transforms of $h, g$. The inner product is defined so that the probability of having a noise realization $n_0(t)$ is $p(n=n_0) \propto \exp[-(n_0|n_0)/2]$. To find the average $\Delta \chi^2$, one then marginalizes the logarithm of the likelihood over many noise realizations (e.g.,~\cite{Cutler:1994ys})
\ba\label{eq:dchisq}
\langle \Delta \chi^2 (\boldsymbol{\theta}) \rangle &\equiv& 2 \langle \log\left[ \mathcal{L} (s | \boldsymbol{\theta}) / \mathcal{L} (s | \bar{\boldsymbol{\theta}}) \right] \rangle  \\
&=&  \left(\,\bar{h} - g \,|\, \bar{h}-g\,\right) \nonumber \\
&=& 4 \int_0^\infty \frac{df}{S_n(f)} \left( H(f,\bar{\boldsymbol{\theta}})^2 + H(f,\boldsymbol{\theta})^2 - 2H(f,\bar{\boldsymbol{\theta}}) H(f,\boldsymbol{\theta}) \cos \left[ \Psi (f, \boldsymbol{\theta}) - \Psi (f, \bar{\boldsymbol{\theta}}) \right] \right)\, \nonumber ,
\ea
where $\mathcal{L} (s | \bar{\boldsymbol{\theta}})$ is the likelihood evaluated at $g=\bar{h}$ with $H$ and $\Psi$ the amplitude and phase of the waveform in the stationary phase approximation. Assuming a Gaussian likelihood, one can interpret $\Delta  \chi$ as the number of ``sigmas" at which the parameter set can be constrained given the noise model.

\subsubsection{Forecasted constraints on $q_i$ and $\lambda$}
\begin{figure}[tbp]
\centering 
\includegraphics[height=0.33\textwidth]{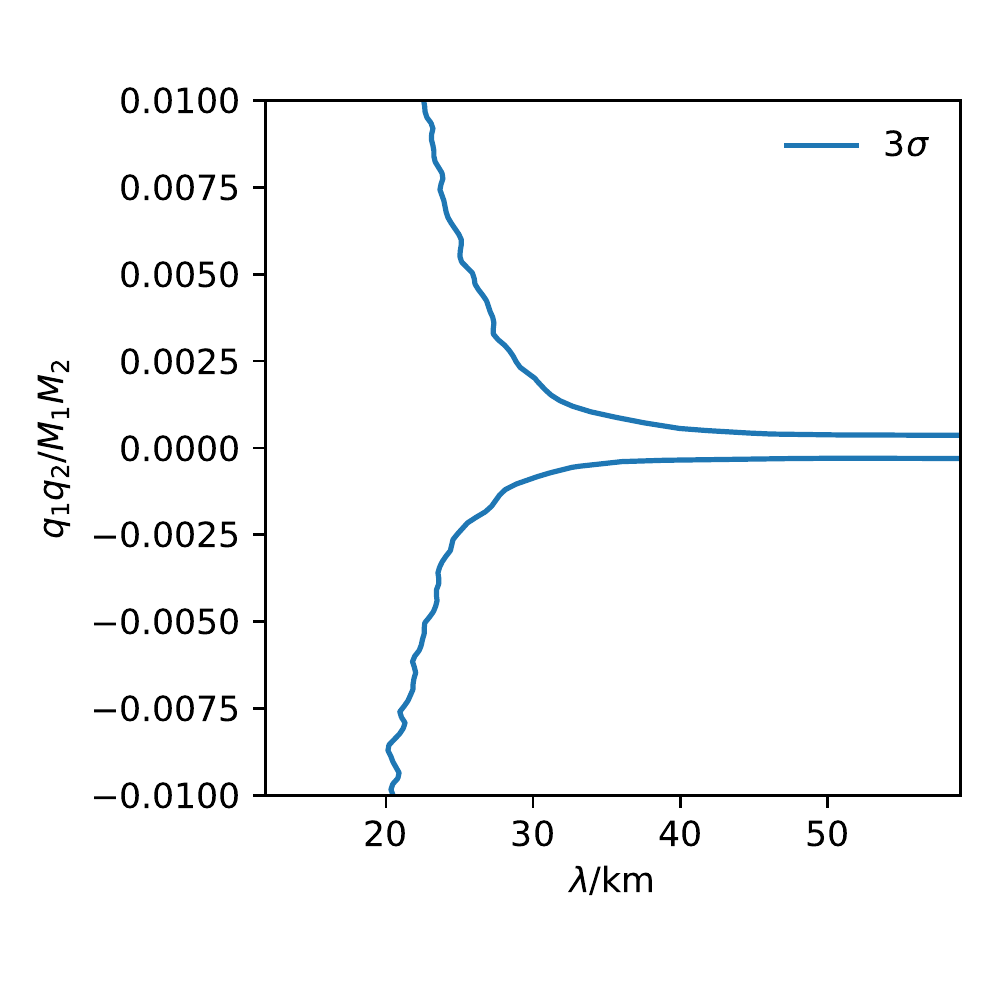} 
\includegraphics[height=0.33\textwidth]{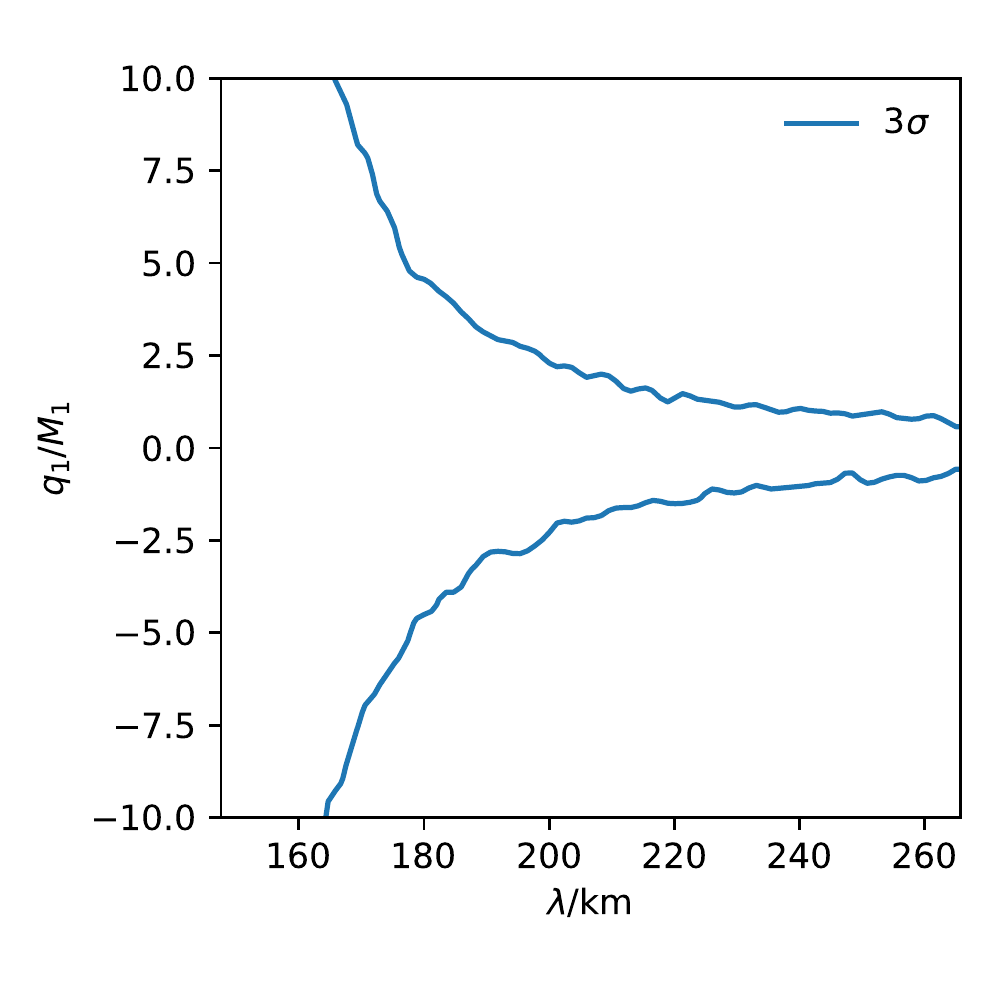} 
\caption{Forecasted marginalized constraints on scalar charge and Compton wavelength. We consider a fiducial model of a binary system with $M_1= 1.2 M_{\odot}$, $M_2 = 1.24 M_{\odot}$ and $D_L=40 \ {\rm Mpc}$ and evolve it in the absence of axions. We sample the likelihood function assuming the noise properties of Advanced LIGO at design sensitivity on the full parameter space. The plot shows the forecasted marginalized 3$\sigma$ constraints on the $(q-\lambda)$ plane (left) and the $(q_1/M_1 - \lambda)$ plane (right).
} \label{fig:mcmc}
\end{figure}
To give an idea of the constraints on the parameters in the axion sector, we consider two fiducial scenarios. In the first scenario, we assume a neutron star binary with masses $M_1= 1.2 M_{\odot}$ and $M_2 = 1.24 M_{\odot}$, evolving at a luminosity distance of $D_L=40 \ {\rm Mpc}$, in pure GR. We also assume the radii of the two neutron stars to be $R^{(1),(2)}_{\rm NS} = 10GM_{1,2}$. We consider a waveform template parameterized by 
\begin{equation}
	\boldsymbol{\theta} = \{{\cal A}, {\cal M}, M, t_c, \phi_c, q_{1,2},  p_{1,2}, \lambda \},
\end{equation}
where ${\cal A} \equiv \sqrt{\frac{5}{24}} \frac{G^{5/6} {\cal M}^{5/6}}{ \pi^{2/3} D_L}$ (the GW amplitude) and ${\cal M}\equiv \mu^{3/5} M^{2/5}$  (the chirp mass). The parameter $t_c$ is the time at which the separation goes to zero in the Newtonian limit and $\phi_c$ is the corresponding phase. Given our assumption of no axion field, the ``true" values of the parameters in the scalar sector are $ q_{1,2} = 0$, $p_{1,2} = 10/16$ and $\lambda = \infty$.

We sample the likelihood function using the {\em emcee} package~\cite{2013PASP..125..306F} on the full 10-dimensional parameter space. We use the forecasted noise curve for Advanced LIGO at design sensitivity (``Design") based on the Zero Det, High Power scenario~\cite{LIGOnoise}. This provides an idea of the noise-limited constraints that could be obtained by Advanced LIGO for a nearby NS-NS inspiral event. In the left panel of Fig.~\ref{fig:mcmc}, we show the marginalized $3\sigma$ forecasted constraints in the $(q - \lambda)$ plane where $q=q_1q_2/M_1M_2$ (see Eq.~\eqref{parameters}). As it can be seen in the plot, there is a degeneracy between $q$ and $\lambda$. Constraints on $q$ become tight as $\lambda$ increases. The GR limit can be achieved as $q \rightarrow 0$ or $\lambda \rightarrow 0$. Hence, it is not a single point in the parameter space. In principle, the 3$\sigma$ constraint contours should approach a non-zero constant as $\lambda$ goes to infinity. However, sampling this infinite ridge in the likelihood function in the large $\lambda$ limit requires a prohibitively large number of samples. Therefore, Fig.~\ref{fig:mcmc} only shows the forecasted constraints in the small $\lambda$ limit.
 
One may expect that the contours approach a constant in $q$ as $\lambda$ goes to infinity (massless axion limit), allowing us to fix this asymptotic constraint by sampling the likelihood function for $\lambda \rightarrow \infty$. We obtain the following $3\sigma$ constraint on q~\footnote{The $3\sigma$ constraint on negative $q$ directly obtained from the MCMC sampling is $-7.2\times10^{-9} < q$, which is tighter than that on positive $q$. This asymmetry, also visible in the left panel of Fig.~\ref{fig:mcmc}, is due to the way we cut the inspiral waveform. In the MCMC sampling, we cut the inspiral waveform at fixed separation. A negative $q$ in the binding energy delays the phase of the waveform, leading to more cycles compared to a positive $q$. Therefore, the constraint on negative $q$ is tighter than on positive $q$. Because the exact constraint is dependent on the cut we choose, we report $\left|q\right| < 6.1\times10^{-8}$ for simplicity. A more complete model of the merger itself is necessary to evaluate the asymmetry between constraints on positive and negative charges.}: 
$$\left|q\right| < 6.1\times10^{-8}.$$
A notable feature about the neutron star solutions discussed in this paper is the induced (``image") charge effects on the axion profile. Given that $8 p > 1$ ($8 p =1$ corresponds to the compaction of black holes, see Eq.~(\ref{eq:p1p2})), q can be tightly constrained due the induced charge effects described by the last term in Eq.~(\ref{Vssimple}), especially in the large $\lambda$ limit where the exponential suppression associated with the Yukawa potential is less important.

As a second scenario, we consider a binary system that consists of a $1.2M_\odot$ neutron star and a $1.24M_\odot$ black hole at $D_L=40 \ {\rm Mpc}$ in pure GR. Note that the parameters were chosen to contrast with the NS-NS case. In reality, astrophysical black holes would have larger masses \cite{Farr:2010tu}. For the same scalar mass, a larger black hole mass would weaken the constraint. For example, for a $4 M_\odot$ black hole, the constraint weakens by a factor of 2. For NS-BH binaries, the only effect the axion has on the inspiral dynamics is through scalar radiation, which can be characterized by $q_1$ and $\lambda$. Thus, the waveform template in this case is parameterized by 
\begin{equation}
	\boldsymbol{\theta} = \{{\cal A}, {\cal M}, M, t_c, \phi_c, q_{1}, \lambda \}.
\end{equation}
We sample the likelihood function using the same method and noise curve as in the first scenario to derive noise-limited constraints that could be obtained by Advanced LIGO for a nearby NS-BH merger event. The marginalized $3\sigma$ constraints in the $(q_1/M_1 - \lambda)$ plane are shown in the right panel of Fig.~\ref{fig:mcmc}. Because scalar radiation can be emitted only if the scalar wave frequency is larger than the mass of scalar field, the constraints on $q_1$ become weaker when $\lambda$ becomes less than the typical wavelength corresponding to $10 Hz$, i.e., the lower bound of the LIGO observational band. Analogously to the first case, we perform the MCMC sampling in the limit of $\lambda \rightarrow \infty$ to resolve the constrains in the limit of large $\lambda$. We find the following $3\sigma$ constraints on q from a NS-BH inspiral event: 
$$\left| q_1/M_1 \right| < 5.7 \times 10^{-4}.$$

Let us briefly compare the constraints from the NS-NS and NS-BH mergers in the $\lambda \rightarrow \infty$ limit considered above. The axion influences the NS-NS merger through an attractive or repulsive scalar force and scalar radiation (note however that scalar radiation is negligible for the case of nearly equal masses chosen here), while it influences the NS-BH merger only through the presence of scalar radiation. For the roughly equal mass binaries that we have considered, we can directly compare the constraint on $q$ for the NS-NS event to $\left| q_1/M_1 \right|^2 < 3.2 \times 10^{-7}$ from the NS-BH event. It can be seen that a stronger constraint can be obtained from the NS-NS event, implying that the scalar force is driving the constraints more than the contribution from scalar radiation.

\subsubsection{Forecasted constraints on the axion parameter space}
\begin{figure}[tbp]
\centering 
\includegraphics[width=0.75\textwidth]{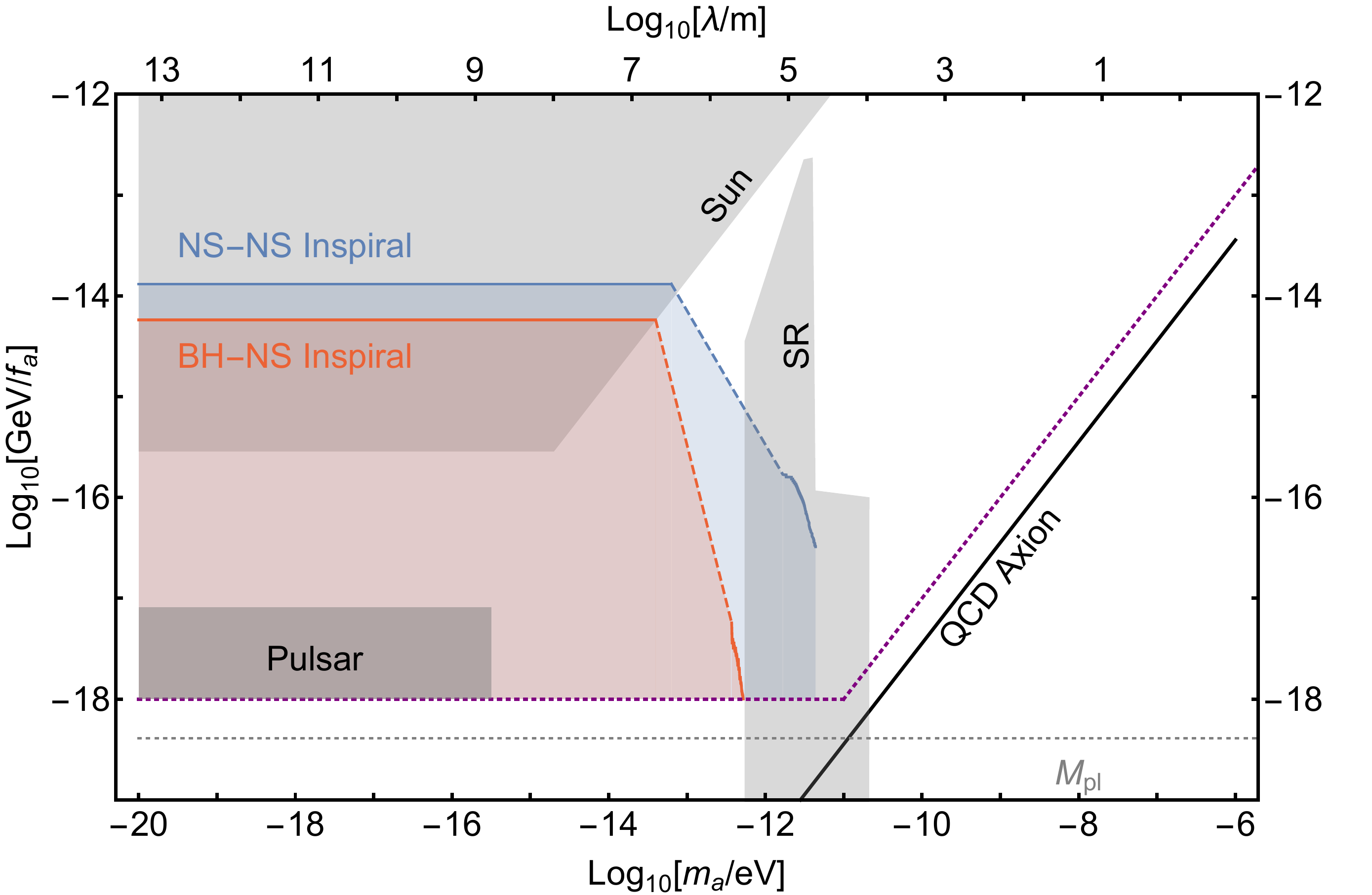} 
\caption{Forecasted marginalized constraints on the axion parameter space. Colored regions below the curves can be constrained by analyzing data of the LIGO detections of NS-NS (blue) and NS-BH (red) mergers. The forecasted constraints are found by performing MCMC sampling in the small $m_a$ limit (the horizontal solid lines) and in the large $m_a$ limit (the curly solid lines). We connect these forecasted constraints by linear interpolation (the dashed lines). Note that the parameters of the NS-BH case were chosen to contrast with the NS-NS case. In reality, astrophysical black holes would have larger masses \cite{Farr:2010tu}. For the same scalar mass, a larger black hole mass would weaken the constraint. For example, for a $4 M_\odot$ black hole, the constraint weakens by a factor of 2. For comparison, we also show the existing constraints (in gray) from direct measurements of the Sun, from measurements of the orbital decay of binary pulsar systems~\cite{Hook:2017psm} and from black hole super-radiance through blackhole spin measurement~\cite{Arvanitaki:2014wva}. A wider range of the axion parameter space can be probed by direct searches of continues wave at LIGO as well as indirect measurement of blackhole spin distribution~\cite{Arvanitaki:2014wva,Arvanitaki:2016qwi,Brito:2017wnc}. The region above the dotted purple line are parameter spaces where an axion profile can be sourced by a neutron star. The black line shows the parameters of a QCD axion, while the dotted gray horizontal line marks the value of the reduced planck scale $M_{\rm pl}$.} \label{fig:mcmcaxion}
\end{figure}

Using Eq.~(\ref{eq:Qtofa}) and Eq.~(\ref{eq:p1p2}), we map $q_i$ to $f_a$, and thus project the constraints in Fig.~\ref{fig:mcmc} to the axion parameter space. The result for the two fiducial binary systems we have studied above are shown in Fig.~\ref{fig:mcmcaxion}. As above, the constraints on $q_i$ are sampled in two regimes, the small $\lambda$ regime and the large $\lambda$ regime. The constraints on the regime in between are found by simple interpolation and are plotted as a dashed line. The NS-NS binary is more constraining than the NS-BH binary due to the stronger constraint on the scalar charge given in the NS-NS case.

Interestingly, our forecasted constraints for Advanced LIGO (the blue shaded region in the plot) are complementary to existing constraints on the axion parameter space, e.g., constraints from direct measurements of the Sun, from measurements of the orbital decay of binary pulsar systems, or of black hole super-radiance (see, Ref.~\cite{Hook:2017psm} for a complete description of these constraints). 
From our analysis, we find that Advanced LIGO has the potential to pin down the axion mass and decay constant within the range
\begin{equation}
	m_a \lesssim 10^{-11} \ {\rm eV}, \quad f_a \gtrsim (10^{14} - 10^{17}) \ {\rm GeV},
\end{equation}
or, in the absence of a detection, to exclude axions with $m_a$ and $f_a$ in this region of parameter space.

Axions with masses and decay constants in the above range are most likely in significant tension with the detected gravitational wave signal from the binary neutron star event GW170817~\cite{TheLIGOScientific:2017qsa}. Having developed the methods and tools to constrain the axion parameter space from a given waveform in this work, it is now possible to perform an analysis of existing and future events. In particular, we would like to apply our analysis to the GW170817~\cite{TheLIGOScientific:2017qsa} data in a follow-up work.

\section{Conclusions and discussion}\label{Sec:conclusion}

In this paper, we have examined the exciting possibility that Advanced LIGO could detect new light scalar particles through their influence on the gravitational waveform produced in NS-NS and NS-BH binary mergers. Employing an EFT approach, we have calculated the first relativistic corrections to the binary orbital dynamics and gravitational waveform in the presence of a light scalar coupled to neutron stars. We use this waveform to forecast the constraints from Advanced LIGO (for an event similar to GW170817) on the parameters of the EFT, which in the scalar sector include the charges of the neutron stars and a relativistic correction corresponding to image-charge effects. This result, summarized in Fig.~\ref{fig:mcmc}, is applicable to theories where a light scalar couples to neutron stars with near gravitational strength. We then specialize to a particularly well-motivated light scalar, the axion. 

If there are in fact axion(s) with mass(es) and decay constant(s) in a region of parameter space where Advanced LIGO has a good sensitivity, then as it can be seen from Fig.~\ref{fig:mcmc}, the parameters of the EFT can be measured with high precision. Such a scenario is most likely under significant pressure from GW170817~\cite{TheLIGOScientific:2017qsa}, however one can speculate about the implications should a detection be imminent. 

The discovery of a new particle using an entirely new observable would, of course, be an incredible development in its own right. One of the first follow-up questions would be: what other physical phenomena could this new particle be related to? Axions with nuclear couplings and masses in this range can potentially solve the strong CP problem of the Standard Model \cite{Hook:2018jle}. In addition, the detected axion could in principle be a dark matter candidate. In the region of parameter space accessible to binary NS mergers, the axion must be produced non-thermally, implying evidence for a non-trivial cosmological history.  

There are a number of avenues for finding corroborating evidence to a detection of axions with LIGO. If the axion is the dark matter, the same nuclear coupling that leads to a force between neutron stars also leads to a time-dependent nuclear electric dipole moment that can be targeted by precision magnetometry~\cite{Graham:2013gfa,Budker:2013hfa} as well as various resonant experiments that look for photon couplings of the axion. Precise knowledge of where to look in parameter space can greatly improve the prospects for detectability using such techniques. A precise knowledge of the masses (and couplings) of the axion significantly narrows the range of axion masses to scan, while the sensitivity to the axion coupling improves as $(m_a/\delta m_a)^{1/4}$ since more time can be allocated to the frequency range where the mass lands (sensitivity scales as $t^{1/4}$). The region of axion parameter space covered by binary NS mergers is also accessible to probes of black hole superradiance, e.g, gaps in the distribution of black hole spin or gravitational waves from rotating axion clouds~\cite{Arvanitaki:2009fg,Arvanitaki:2010sy}.

The axion would also provide an interesting additional probe of the structure of the merging neutron stars. From Eq.~(\ref{AxionForce}), the scalar charge of the individual neutron stars is dependent upon the compaction (recall that the compaction is defined as $GM/R$) and the decay constant $f_a$. Also note that from Eq.~(\ref{eq:p1p2}), the EFT parameter $p$ is sensitive to the compaction of the neutron stars. For an event with sufficiently high SNR, the axion force therefore provides a new way to constrain the compaction of neutron stars. With future detectors, it may also be possible to use the post-merger waveform associated with the hyper-massive neutron star resulting from the merger event to provide further knowledge on both the properties of the axion and nuclear equation of state~\cite{Sagunski:2017nzb}. We leave further investigation of the post-merger signal for future work.

In the absence of a detection, it is possible to set stringent constrains on the region of parameter shown in Fig.~\ref{fig:mcmcaxion}, for axions that possess a nuclear coupling. It is not necessary for all such particles to possess a nuclear coupling. Nevertheless, the lack of a detection would imply that laboratory experiments relying on such couplings, such as CASPEr-Electric~\cite{Budker:2013hfa}, should also fail to make a detection over the same region of parameter space. Knowing where not to look could be useful in guiding such searches. The effects associated with superradiance could in principle be found even in the lack of a detection from binary neutron stars. In this case, one would strongly constrain the nuclear coupling of the axion, and potentially the QCD axion (the only target for laboratory experiments looking for an axion through its nuclear couplings).

Let us also briefly comment on the possibilities with future gravitational wave detectors. For the purpose of constraining light scalars from neutron star mergers, Advanced LIGO is limited by its overall sensitivity and its frequency coverage. An increase in sensitivity over the Advanced LIGO band, as would be provided by third generation gravitational wave detectors such as Einstein Telescope~\cite{Sathyaprakash:2012jk}, would yield a number of advances. The projected constraints on scalar charge would become tighter as the SNR per event would be higher. Greater detection rates would allow for a joint analysis of many events (e.g. ``stacking'') that could provide stronger projected constraints than individual events. In addition, greater sensitivity at high frequencies could provide access to the end-stages of the inspiral and the ring-down of the hyper-massive neutron star or black hole that can form as a result of the merger. This would provide new information about the scalar sector through additional relativistic corrections, and through effects on the structure and evolution of post-merger objects (for example, as explored in Ref.~\cite{Sagunski:2017nzb}). A space mission such as LISA~\cite{AmaroSeoane:2012km} will provide sensitivity at lower frequencies. For individual events, this would provide access to scalars with a lower mass as the binary evolution could be tracked at larger separation. In addition, the projected reach on the charge dipole of the binary would improve since orbital energy-loss due to scalar radiation is more important at lower frequencies. Finally, it will be possible to observe the merger of white dwarfs (either individually or as a stochastic background), which would allow one to examine the nature of the coupling between axions and compact objects. In particular, it would be interesting to examine the density-dependent coupling invoked in the axion model we have studied here.

Beyond axions, our results are applicable to more general scalar tensor theories. Previous literature on massive scalar tensor theories has mainly focused on Brans-Dicke theory~\cite{Alsing:2011er}, including extreme-mass ratio binaries~\cite{Yunes:2011aa} and NS-BH binaries~\cite{Berti:2012bp} as well as NS-NS systems exhibiting spontaneous scalarization~\cite{Ramazanoglu:2016kul,Alby:2017dzl}. The present work extends these studies to include all relevant couplings to 1PN order for massive scalar tensor theory in the Einstein frame. In particular, we have highlighted the importance of the image-charge effect. Future studies could explore the relevant matching conditions between the EFT and various scalar tensor theories.

In summary, the observation of binary neutron star mergers provides a novel opportunity to search for new light scalar particles, including axions. The waveforms presented in this paper, and the forecasted constraints, provide the technical basis and proof-of-concept necessary to proceed with an analysis of data from existing and future events. In particular, we hope to perform an analysis using data from the existing event GW170817 in future work. The results of such an analysis will greatly inform other observational and laboratory efforts to search for light scalars, and provide constraints over an extensive and well-motivated region of parameter space for axions.

\acknowledgments
We thank Anson Hook for collaboration in the early stages of the project. We thank Asimina Arvanitaki, Masha Baryakhtar, Anson Hook, Adrien Kuntz, Robert Lasenby, Luis Lehner, Federico Piazza, and Huan Yang for helpful discussions. MCJ is supported by the National Science and Engineering Research
Council through a Discovery grant. MS is partially supported by the STFC grant ST/L000326/1. This research was supported in part by Perimeter
Institute for Theoretical Physics. Research at Perimeter Institute is supported
by the Government of Canada through the Department of Innovation, Science and
Economic Development Canada and by the Province of Ontario through the Ministry
of Research, Innovation and Science.

\appendix

\section{Degeneracy with higher PN corrections}\label{App:A}

\begin{figure}[tbp]
\centering 
\includegraphics[height=0.4\textwidth]{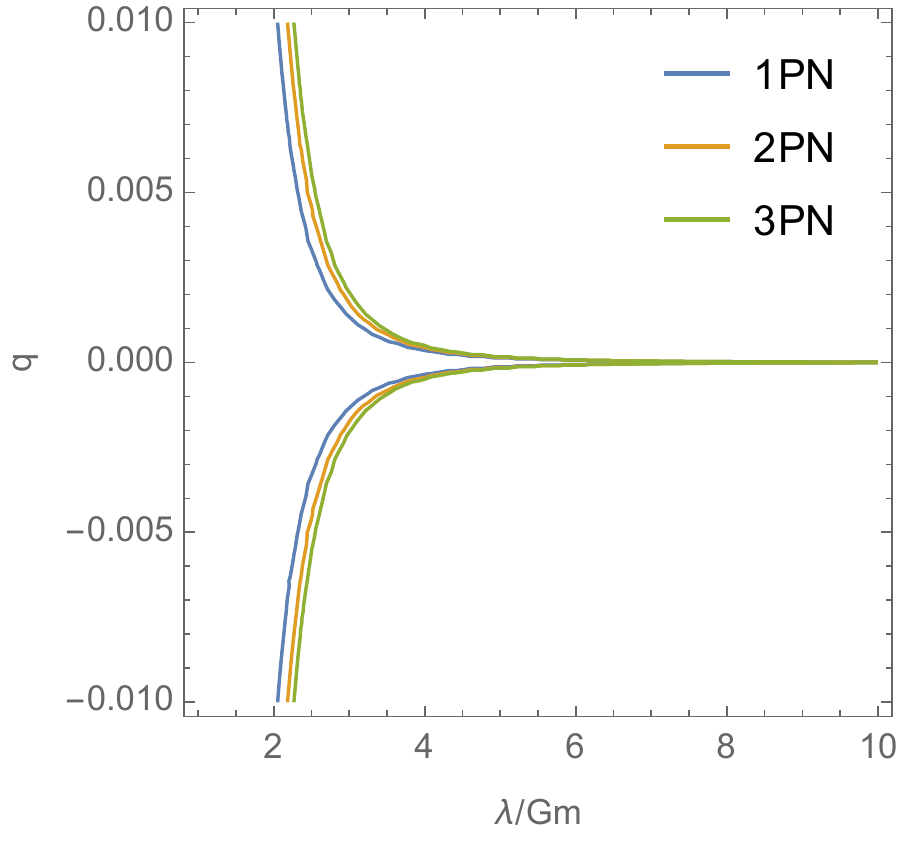} 
\caption{Phase difference caused by the scalar with different PN GR expressions. We consider a binary of masses $1.2 M_{\odot}$ and $1.24 M_{\odot}$, and calculate the total phase difference (integrated from 10Hz to 1000Hz) caused by the scalar. The contours show the total phases between the cases with and without scalar differ by 1. $\lambda$ is in units of the total inverse mass.} \label{fig:PNeffects}
\end{figure}

The waveform obtained above considers only 1PN corrections to GR. In principle, one can improve the waveform by simply replacing the 1PN expressions of the gravity sector with higher PN expressions. In this section, we estimate the impact of including higher PN corrections in the gravity sector on our constraints in the scalar sector. The constraints on the scalar sector can be characterized by the phase difference caused by the axion field. Specifically, we consider a binary system composed by $1.2 M_{\odot}$ and $1.24 M_{\odot}$ masses and assume the two stars carry the same scalar charge. We calculate the total phases, $\Psi_{\rm GR}$ and $\Psi_{\rm s}$, by integrating the phase over (10 -1000) Hz in the cases with and without the scalar. The constraints on the scalar sector can be characterized by the differences of total phases, $\Delta\Psi = \left|\Psi_{\rm s} - \Psi_{\rm GR}\right|$. From Eq.~\ref{eq:dchisq}, it can be seen that $\Delta \chi^2$ can be significant only once the phase difference is order one. We calculate $\Delta\Psi$ using different PN order expressions in gravity sector, and the plot the contours of $\Delta\Psi=1$ in Fig.~\ref{fig:PNeffects}. We find that the region of parameter space over which the phase difference is order one, as shown in Fig.~\ref{fig:PNeffects}, does not change significantly when including higher PN terms in the gravity sector, especially for the parameter range we are interested in. It is therefore justified to just use the 1PN correction in order to forecast constraints on the scalar EFT parameters. 

\bibliography{references}

\end{document}